\documentclass[%
 reprint,
superscriptaddress,
longbibliography,
 amsmath,amssymb,
 prl,
floatfix,
]{revtex4-1}

\newcommand{\beginsupplement}{%
        \setcounter{table}{0}
        \renewcommand{\thetable}{S\arabic{table}}%
        \setcounter{figure}{0}
        \renewcommand{\thefigure}{S\arabic{figure}}%
        \setcounter{equation}{0}
        \def\theequation{S\arabic{equation}}
     }

\usepackage{graphicx}
\usepackage{dcolumn}
\usepackage{bm}
\usepackage[colorlinks=true,
            linkcolor=blue,
            urlcolor=blue,
            citecolor=blue]{hyperref}

\usepackage{algorithm}
\usepackage{algpseudocode}

\begin{document}

\title{Phenotypes of vascular flow networks}

\author{Henrik Ronellenfitsch}
\email{henrikr@mit.edu}

\affiliation{Department of Mathematics, Massachusetts Institute of Technology, Cambridge, MA 02139, USA}

\affiliation{Department of Physics and Astronomy, University of Pennsylvania, Philadelphia, PA 19104, USA}
\author{Eleni Katifori}%
\email{katifori@sas.upenn.edu}

\affiliation{Department of Physics and Astronomy, University of Pennsylvania, Philadelphia, PA 19104, USA}

\date{\today}

\begin{abstract}
Complex distribution networks are pervasive in biology. Examples include nutrient transport in the slime
mold \emph{Physarum polycephalum} as well as mammalian and plant venation.
Adaptive rules are believed to guide development of these networks and lead to a reticulate,
hierarchically nested topology that is both efficient and resilient against perturbations. However,
as of yet no mechanism is known that can generate such networks on all scales.
We show how hierarchically organized reticulation can be constructed and maintained through spatially
correlated load fluctuations on a particular length scale.
We demonstrate that the network topologies generated represent a trade-off between optimizing transport efficiency, construction cost, and
damage robustness and identify the Pareto-efficient front that evolution is
expected to favor and select for. We show that the typical fluctuation length scale controls the position of the networks on
the Pareto front and thus on the spectrum of venation phenotypes.
\end{abstract}

\maketitle
Complex life would be inconceivable without biological fluid distribution networks such as animal vasculature,
plant xylem and phloem, the network of fungal mycelia or the protoplasmic veins of \emph{Physarum polycephalum}.
These networks distribute oxygen and nutrients, remove waste and serve as long range
communication pathways.
In mammals, the vast spectrum of venation network
phenotypes ranges from
predominantly tree-like networks such as the large veins and arteries that service entire organs to
highly reticulate capillaries within the organs such as in the brain or the liver.
In plants, leaf network phenotypic variability even within a single organism can be large, but typically the
hierarchical structure and reticulation are roughly conserved.
However, within a single family
there can be considerable variation~\cite{Ronellenfitsch2015b}.
It is therefore natural to ask whether there might be a single developmental mechanism at play that
can generate and interpolate between the different archetypes on this
phenotypic spectrum of vascular networks. Then,
evolution would only need to select for a few parameters in order to tune
the network phenotype for
its function. Here, we theoretically identify fluctuations during
development as such a mechanism, and pinpoint networks on a Pareto
front possessing optimal trade-offs between hydraulic efficiency, damage resilience,
and cost, as evolutionarily desirable.

\begin{figure}[t]
\centering
\includegraphics[width=.98\columnwidth]{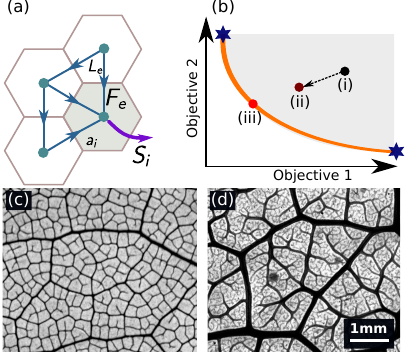}
\caption{(a) Network model of liquid transport. Edges $e$ of length $L_e$ carry currents
$F_e$. At each node $i$, a net current $S_i$ is drawn from the network. The net current $S_i$ models
local sources and sinks.
(b) The Pareto front (orange) is the set of points out of all possible
phenotypes (gray) for which performance can not be improved at both objectives
simultaneously. For any point not on the Pareto front, e.g., (i), a different point can be
found, e.g., (ii), that has better performance at both objectives. For a point on the
Pareto front, like (iii), this is not possible.
The endpoints of the Pareto front (stars) are functional archetypes.
(c) Leaf veins of \emph{Acer platanoides}
near the ``reticulate archetype''
identified in this paper.
(d) Leaf veins of \emph{Protium dawsonii} show many freely ending
veinlets, similar to what is found near the ``tree archetype'' identified
in this paper.
\label{fig:figure1}}
\end{figure}

Many frequently competing factors influence which particular phenotypes are favored by natural
selection. Therefore, it is to be expected that the eventual physical
form of an organism is shaped by trade-offs between different requirements.
Pareto optimality identifies those phenotypes that strike optimal trade-offs
between objectives: The Pareto front is the subset of phenotypes
where performance at one objective can not be increased without decreasing performance
at another \{Fig.~\ref{fig:figure1}~(b), Ref.~\cite{Miettinen1999}\}.
One can assume that the phenotypes observed in nature are found approximately
on some relevant Pareto front because any other trade-off could be improved upon and is
therefore evolutionarily selected against, given otherwise fixed conditions~\cite{Shoval2012}.

In plants, where a well preserved fossil record of the venation exists, the fast transitions between reticulate and non-reticulate patterns over evolutionary time are evidence for an easily tunable mechanism generating vascular phenotypes ~\cite{Givnish2005,Blonder2016}.
These transitions can also be effected artificially by
single gene knockouts~\cite{Steynen2003,Carland2009} or small changes in phytohormone concentrations~\cite{Berleth2000}.
In the case of animals, often the positions and dimensions of the largest vessels (such as the aorta)
are genetically predetermined and fixed.
However, smaller vessels are too numerous to be efficiently genetically
encoded and are believed to develop in a self-organized fashion~\cite{LeNoble2005,Kurz2001,Nguyen2006}.
The abstract mechanisms governing self-organization of vasculature
in plants and animals appear to be universal~\cite{Ronellenfitsch2016}.
For instance, in plant leaves, auxin canalization, involving flow of a chemical morphogen,
is believed to guide development of the network pattern \{Refs.~\cite{Smith2009,Scarpella2006,Verna2015,Feugier2005,Feller2015}, Fig.~\ref{fig:figure1}~(c,d)\}
and in animal vasculature, vessels
respond to wall shear stress~\cite{Eichmann2005,Hu2012,Kurz2001,Scianna2013,Hacking1996}.
Generically, these mechanisms involve a process that is able to remodel an initial mesh of veins
according to the flow of blood (in animals), or cells connected by carrier proteins according to a morphogen (in plants).
If the flow is large, vessels adapt by increasing their diameter;
unused connections die out. This process has been observed directly in animals~\cite{Chen2012}
and indirectly in plants~\cite{Marcos2014}.

Common to the vascular network development of both plants and animals, the dynamics of the
hydraulic vessel conductivities $K_e$ can be
modeled by an equation of the form~\cite{Hu2013,Ronellenfitsch2016,Hacking1996,Rolland-Lagan2005,VanBerkel2013},
\begin{align}
\frac{d K_e}{dt} = a \frac{{(F_e^2)}^{\beta}}{{K_e}^{\alpha-1}} - b K_e + c \,e^{-rt},
\label{eq:adapt-family-mod2}
\end{align}
where $a$, $b$, $c$ and $r$ are non-negative adaptation parameters and $\alpha\geq 1$, $\beta > 0$. Often, $\alpha=1$ and $c=0$.
The dynamical steady states then correspond to different network topologies.

\begin{figure*}
\centering
\includegraphics[width=.98\textwidth]{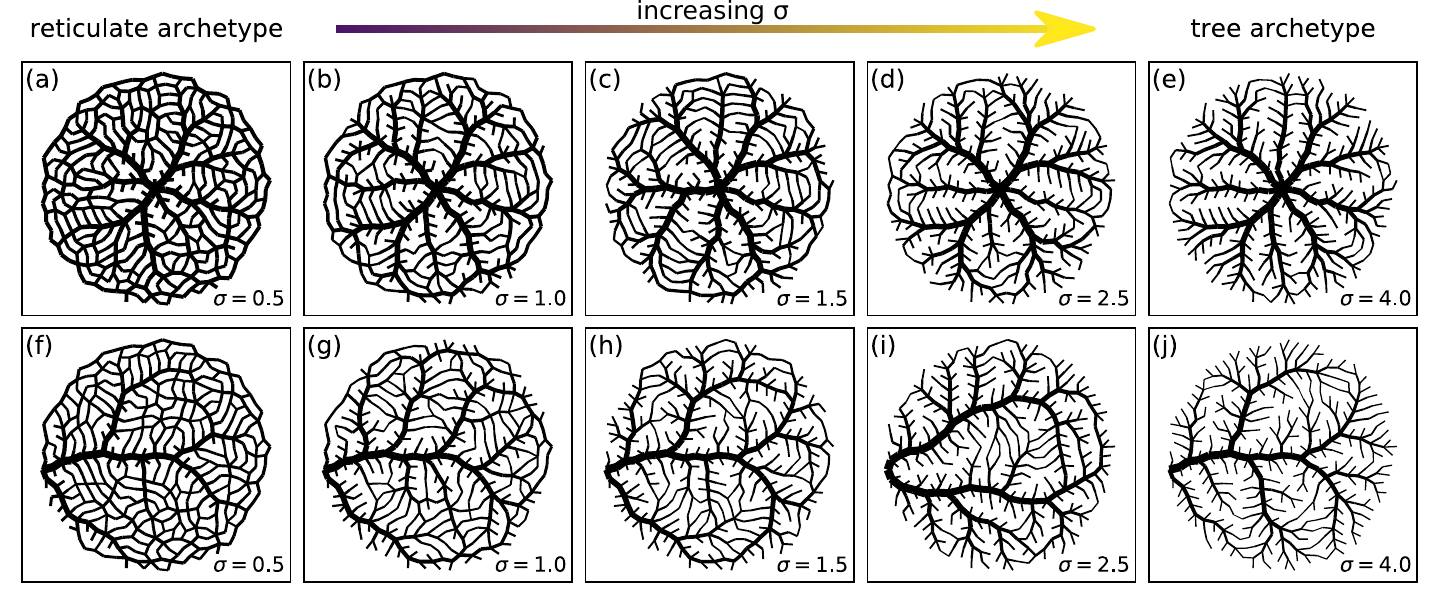}
\caption{The variety of network phenotypes that can be produced with a locally adaptive fluctuating load model.
All examples lie on the Pareto front of efficient networks (Fig.~\ref{fig:pareto}),
thus representing different trade-offs between baseline power dissipation, cost, and damage robustness.
The number of loops and thus damage robustness increases to the right, the value of $\sigma$ increases
from 0.5 to 4.0 to the right.
The Pareto front corresponds to the whole spectrum of ``natural'' reticulate networks, from
highly hierarchical trees, fragile but cheap, to highly robust reticulate, expensive networks.
(a)--(e) The inlet is at the center. 
(f)--(j) The inlet is at the left side. 
\label{fig:variety}}
\end{figure*}

Equation~\eqref{eq:adapt-family-mod2} describes a local positive feedback mechanism.
Conductivities $K_e$ grow as controlled by the magnitude of $a$ when the
current $F_e$ through their vessel is large, and they
decay on a characteristic time scale $b^{-1}$ when it is small.
The parameter $c$ may be interpreted as the presence of some growth factor
such as VEGF in the case of mammalian vasculature or background production of auxin
transporting proteins in the case of plant leaves~\cite{Rolland-Lagan2005}.
Potential flow is assumed throughout \{Fig.~\ref{fig:figure1}~(a), Supplemental Material~\footnote{See Supplemental Material [url], which includes
Refs.~\cite{Blakeslee2005,Kramer2006}.}\}. An explicit time-dependence may exist during development, for instance due to growth of
the surrounding tissue, or gradual
depletion or degradation of the growth factor
over a time scale $r^{-1}$~\cite{Ronellenfitsch2016}.

The generic dynamics of Eq.~\eqref{eq:adapt-family-mod2}
is characterized by two phases. First, the background production term dominates and
produces a homogeneous network.
Then, as background production becomes increasingly
suppressed due to the exponential decay term, vascular adaptation takes over,
generating veins in a hierarchical fashion: thick, main veins first and successively
thinner veins later while pruning unused connections, comparable to vascular plexus development~\cite{Fleury2007,Eichmann2005,Fruttiger2007,Chen2012}.
The competition between background production
and adaptation leads to hierarchically ordered steady-state
networks~\cite{Ronellenfitsch2016}, which are
always
topological trees~\cite{Bernot2009,Banavar2000}.
While non-hierarchical reticulation can
be achieved by postulating new chemicals~\cite{Feugier2006}, we now introduce a model
of adaptation to fluctuating load that can produce hierarchical reticulation.
Such load fluctuations are common in animals (for instance Ref.~\cite{Drew2011})
and recent work points toward their existence in plants during development as well~\cite{Marcos2014}.

Assuming that the time scale on which fluctuations occur is much smaller than that of adaptation
and that fluctuations are characterized by approximately static states
between which the system switches quickly,
we replace the squared currents in Eq.~\eqref{eq:adapt-family-mod2} by a fluctuation average~\cite{Hu2012,Hu2013,Corson2010,Katifori2010,Ronellenfitsch2018a,Graewer2015,Martens2017},
\begin{align}
F_e^2 \rightarrow \langle F_e^2 \rangle = \frac{1}{N}\sum_{\text{state } i} \left(F^{(i)}_e\right)^2.
\end{align}
Here, the vector of fluctuating states $\mathbf F^{(i)} = ( F^{(i)}_e )$ represents the flows in the network for a particular vector of source terms
$\mathbf S^{(i)} = ( S^{(i)}_j )$, and the summation performs an ensemble average for a given set of fluctuating states.
Then, dynamical steady states can correspond to minima of
optimization models~\cite{Hu2013,Corson2010,Katifori2010}.

We generalize these approaches to include
collectively produced fluctuations by using
the sources,
\begin{align}
\frac{S^{(i)}_j}{\hat S} = \delta_{j0} - (1 - \delta_{j0})\, f\left(\frac{\|\mathbf x_j - \mathbf x_i\|}{\sigma} \right),
\label{eq:fluct-new}
\end{align}
where $\mathbf x_i$ is the position of node $i$, $\sigma$ is the scale over which the source strength
varies, and
$\sum_j S^{(i)}_j = 0$. The total in- and outflow is $\hat S$.
In the rest of this paper we consider Gaussian sources ($f(x) \sim e^{-x^2/2}$).
Other $f(x)$ lead to qualitatively similar results (Supplemental Material~\cite{Note1}).
Uncorrelated fluctuations are obtained as $\sigma\to 0$ and
lead to reticulation, but not to significant hierarchical ordering, similar to Fig.~\ref{fig:variety}~(a,f).

We numerically solve a dimensionless
form of Eq.~\eqref{eq:adapt-family-mod2},
\begin{align}
\frac{d \tilde K_e}{d\tilde t} = \langle \tilde F_e^2\rangle ^{\beta} - \tilde K_e + \kappa\, e^{-\tilde t/\rho},
\label{eq:dimensionless}
\end{align}
where the tilde denotes dimensionless
quantities~(Supplemental Material~\cite{Note1}).
Following Ref.~\cite{Ronellenfitsch2016} we set $\alpha = 1$, with other values leading to similar conclusions. The control
parameters are the dimensionless background strength $\kappa = (c/a) \hat S^{-2\beta}$, the decay timescale $\rho=b/r$, and the
fluctuation scale $\sigma$.
We further fix the nonlinearity at $\beta = 2/3$, which leads to the same
steady-state networks as shear-stress adaptation~\cite{Hu2013}. This value also
corresponds to a total network volume constraint
~\cite{Ronellenfitsch2016,Katifori2010}.
All networks start from the same disordered mesh with 445 nodes and 1255 edges.
We either place a single inlet
at the center of the network, similar to
the retina, or at the boundary, similar to a leaf.
The conductivities are initialized with random positive numbers,
and the scale parameter $\sigma$ is measured in units of the
mean edge length $\hat L$.
\begin{figure*}[t]
\includegraphics[width=\textwidth]{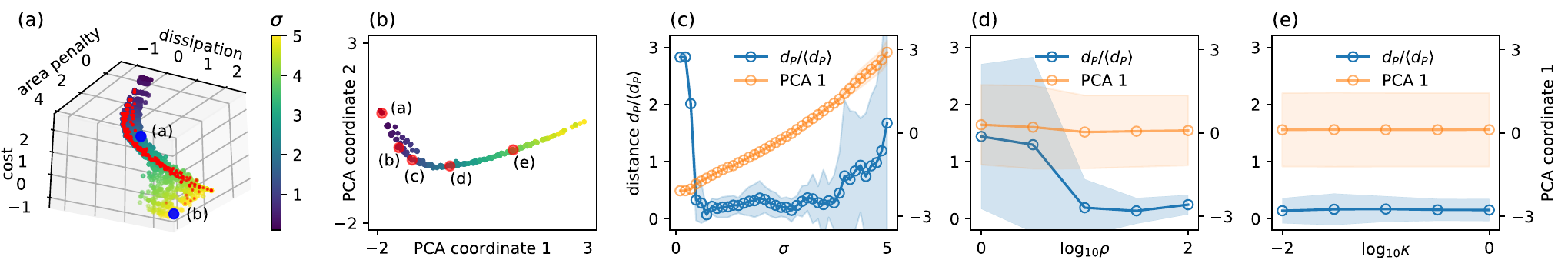}
\caption{Geometry of the Pareto front of adaptive distribution networks. We plot the phenotypic space
of networks obtained from parameter values $\rho \in \{1, 10, 100\}$, $\kappa \in \{ 1, 0.1, 0.01\}$,
$\sigma \in [0.1, 5]$, $\alpha=1, \beta=2/3$ as an example of the phenotypic space that
can be reproduced using the model.
We calculate the Pareto front for simultaneous minimization of power dissipation, network
cost, and percolation penalty.
The data was scaled to zero mean and unit variance in each objective.
(a) The data set,
colors indicate the value of $\sigma$. The Pareto front is in red, and the non-Pareto networks from Fig.~\ref{fig:figure4} are in blue.
(b) Principal component analysis (PCA) embedding of the Pareto front
from (a). $91\%$ of the variance is encoded in the first PCA coordinate, suggesting
that the front is approximately one-dimensional. The first
PCA coordinate (PCA 1) is approximately parametrizes the Pareto front.
Red points correspond to the networks from Fig.~\ref{fig:variety} (a--e).
(c) For all combinations $(\rho,\kappa,\sigma)$ on the Pareto front $P$
we hold $\rho$ and $\kappa$ fixed and vary $\sigma$. For a wide
range of $\sigma$, the average distance $d_P(x) = \min_{p\in P} \|x-p \|$
from the Pareto front is well below the mean $\langle d_P \rangle$,
suggesting that phenotypes remain close to the Pareto front (blue curve,
shaded region is one standard deviation over combinations
of $\rho, \kappa$).
Varying $\sigma$ moves linearly along the Pareto front
parametrized by PCA 1 (orange curve).
Thus, $\sigma$ approximately parametrizes the Pareto front.
Similarly varying $\rho$ (d) or $\kappa$ (e) while holding the other
parameters fixed may lead to phenotypes close to the Pareto front
for large $\rho$ and all $\kappa$, but the position on the
front PCA 1 is random.
Thus, $\rho$ and $\kappa$ can not be used to parametrize the Pareto front.
\label{fig:pareto}}
\end{figure*}

The interplay between background and
decay parameters, fluctuation scale, and
boundary conditions leads to a whole spectrum of networks,
many of them qualitatively resembling
the networks found in dicot and fern leaves, or the vasculature of the retina or the brain.
They appear to reproduce well the hierarchical structure
seen in real modern plants and animals~(Fig.~\ref{fig:variety}).
Reticulation in particular is controlled by the
fluctuation scale $\sigma$.
For small $\sigma \ll \hat L$, the steady state networks
are highly reticulate, similar to those obtained in
Refs.~\cite{Katifori2010,Hu2013}, and have little hierarchy~[Fig.~\ref{fig:variety}~(a),(b),(f),(g)].
As $\sigma$ becomes comparable to or greater than $\hat L$,
the networks gradually lose reticulation and
gain hierarchical structure, independent of the
chosen inlet position~[Fig.~\ref{fig:variety}~(c)--(e), (h)--(j)].
Intuitively, different large-scale sources $\mathbf{S}^{(i)}$
centered at nearby nodes overlap almost completely, and effectively
act as a single state. Thus, the average is over only a few effective,
large-scale sources, which leads to fewer effective fluctuations and
therefore less reticulation. We develop a unified framework for
arbitrary fluctuating sources
by noting that the average flow can be rewritten as the weighted mean (Supplemental Material~\cite{Note1}),
\begin{align}
    \langle F_e^2 \rangle = \frac{1}{N}\sum_{i} \left(F^{(i)}_e\right)^2
    = \sum_j \rho_j \left(R^{(j)}_e\right)^2,
\end{align}
where the $\rho_j$ are the eigenvalues of the covariance matrix
$\frac{1}{N}\sum_k\mathbf{S}^{(k)} (\mathbf{S}^{(k)})^\top$, and the $R_e^{(j)}$ are the
flows induced by the associated eigenvectors as sources.
For values of $\sigma \gg \hat L$,
the collective sources themselves become highly correlated to each other,
and the source covariance matrix is characterized by only a few
dominant eigenvalues, with the vast majority negligibly small, independent of the specific form of $f(x)$ (Supplemental
Material~\cite{Note1}).
Armed with this model, we proceed to ask which of the
network topologies it can produce may be favored by natural selection.
We specialize to a single inlet at the
center, with other inlet positions leading to
qualitatively similar results (Supplemental Material~\cite{Note1}).

Hydraulic efficiency, low cost, and robustness are important but competing requirements, such that we expect
that natural selection strikes a trade-off between them.
As a measure of network efficiency, we consider the hydraulic power dissipation calculated under non-fluctuating conditions,
$E = \sum_e L_e F_e^2/K_e$,
where the flows are computed for a single inlet and uniform sinks.
The rationale is that during nominal operation, fluctuations are expected to be small,
with large fluctuations to be expected during development.
Next, the network cost,
$C = \sum_e L_e K_e^\gamma$,
where $\gamma < 1$ models an economy of scale,
measures the amount of material investment
that goes into constructing the network.
This should be minimized by any organism that efficiently uses its resources.
We set $\gamma=1/2$, which
corresponds to a cost proportional to the total vessel volume,
or equivalently, total material used to construct the network.
Finally, we consider a
percolation penalty as a measure of network robustness,
quantifying the cost of losing part of the vasculature to
damage. We choose the expected fraction of perfused area lost
upon removing an edge,
$\hat A = (1/N_e)\sum_e A_e/A_{\text{tot}}$,
where $A_e$ is the area of the network that becomes disconnected from the source
upon removal of edge $e$, $A_\text{tot}$ is the total area of the network,
and $N_e$ is the number of edges.
Efficient network phenotypes must minimize
the cost $C$, the power dissipation $E$, and the
percolation penalty $A$.

\begin{figure}[b]
\centering
\includegraphics[width=.85\columnwidth]{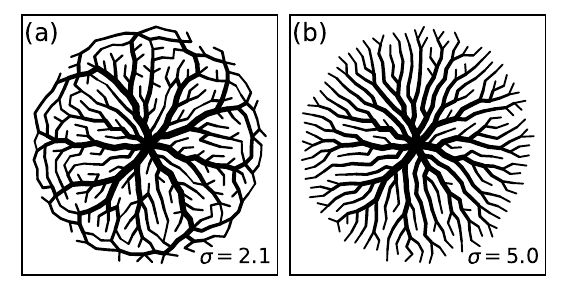}
\caption{Network phenotypes not lying on the Pareto front show
less hierarchical organization for the same amount of reticulation than their Pareto optimal counterparts. In both networks, $\kappa = 1.0$, $\rho=1.0$. (a) $E=6.0$, $A=0.48$, $C=47.0$. (b)
$E=10.8$, $A=7.73$, $C=32.8$.
\label{fig:figure4}}
\end{figure}

Observations of real networks, for instance in leaves, reveal that many treelike components
exist and that they are important for transport~\cite{Fiorin2015}.
This means that although the percolation penalty is minimized,
it is not expected to be perfectly zero.
Except for very small $\sigma \ll \hat L$ and very large $\sigma \gg \hat L$, network phenotypes
obtained from our model generically
exhibit these small treelike components within loops~[Figs.~\ref{fig:variety}, \ref{fig:figure4}].

We scanned a portion of the
parameter space and computed the three network measures for a data set of steady states
of the adaptation dynamics.
The steady state networks form a dense cloud in the space of
network measures~[Fig.~\ref{fig:pareto}~(a)].
Computing the Pareto front using the algorithm from Ref.~\cite{Geilen2007} and analyzing its geometry using Principal Component Analysis (PCA)
reveals an approximately one-dimensional line of points
\{Fig.~\ref{fig:pareto} (b), Supplemental Material~\cite{Note1}\}.
Fixing $\rho$ and $\kappa$, the parameter $\sigma$
approximately parametrizes networks on the Pareto
front~[Fig.~\ref{fig:pareto} (c--e)], such that
$\sigma$ can be used to tune optimal trade-offs between the
three objectives.
The endpoints of the Pareto front correspond to functional
archetypes~\cite{Shoval2012},
on one end low-cost, fragile and non-reticulate,
high dissipation networks ($\sigma \gg \hat L$, tree archetype),
and on the other end high-cost, robust and fully reticulate,
low dissipation networks ($\sigma \ll \hat L$, reticulate archetype)
[Figs.~\ref{fig:variety}, \ref{fig:pareto}].
For small $\sigma \ll \hat L$,
most networks lie
close to the front, whereas for large
$\sigma \gg \hat L$, there is greater variability, and
many networks lie far from the front [Fig.~\ref{fig:pareto} (a),(b)].
Defining a distance $d_P(x) = \min_{p \in P} \|x-p\|$
from the Pareto front $P$ and rescaling all network measures
to have unit variance and mean zero so as to bring them
to the same scale, the mean distance from the
front is $\langle d_P \rangle \approx 0.24$. The Pareto front
comprises $14\%$ of all networks. From the remaining ones,
$70\%$ lie closer than average to the front
and $30\%$ lie further than average from the front.
Tuning $\kappa$ by itself without fixing
the other parameters has little effect on the distance of
networks from the Pareto front. However, $\rho\gtrsim 10$ or
$0.5\lesssim \sigma \lesssim 3$ can generically
drive the network phenotypes close to the front
(Supplemental Material~\cite{Note1}).
Non-Pareto optimal phenotypes often show branching with parallel instead of
roughly perpendicular veins (Fig.~\ref{fig:figure4}).
Open, non-hierarchical venation patterns similar to those of some
networks off the Pareto front can be found in
in the leaves of the evolutionarily archaic \emph{Ginkgo biloba} tree
\{Fig.~\ref{fig:figure4}~(b), Refs.~\cite{Zhou2003,Dorken2014}\}.

We have shown that a simple, easily tunable mechanism is able to
produce an entire
spectrum of phenotypic variation in vascular networks.
The shape of networks on this spectrum can be rationalized by
the interplay between flow fluctuations affecting developmental
processes, and natural selection of parameters that lead
to phenotypes on a Pareto front of optimal trade-offs between
efficiency, cost, and resilience.
The networks on the Pareto front are reminiscent of modern
natural leaf or animal vasculature, suggesting that
natural networks may be subject to the trade-offs we consider.
Networks away from the Pareto front generically
exhibit less hierarchical organization
and less resemblance to modern plants and animals.
Out of the three control parameters of our model, only the
fluctuation scale is highly correlated to the position on the Pareto front and thus to the position on the spectrum of vascular networks.
This could allow natural selection to more easily adjust for a given needed functionality, but also to re-use the same genetic pathway to construct networks with different functionality in the same organism.
Beyond biology, engineered transport
networks such as electrical power grids are often
subject to similar trade-offs, such that we expect
that our analysis will be useful here as well.

\acknowledgments{E.K. acknowledges support by NSF Award PHY-1554887, IOS-1856587, the University of Pennsylvania Materials Research Science and Engineering Center (MRSEC) through award DMR-1720530, the University of Pennsylvania CEMB through award CMMI-1548571, and the Simons Foundation through award 568888 and the Burroughs Welcome Career Award.}

\newpage
\onecolumngrid
\appendix

\begin{center}
\Large\textbf{Supplemental Material}
\end{center}

\beginsupplement
\section{Potential flow in vascular networks}
Here, we describe a general framework capable of describing potential-driven flow of some
quantity through a network that dynamically adapts its conductivities.
Each node is taken to represent a unit of some subdivision of the underlying
tissue, a basin that is fed by that node,
with edges representing the flow between these basins either through vessels or via a
facilitated diffusion process.

The current $F_e$ through each edge $e$ connecting adjacent units $i$ and $j$
is given by $F_e = K_e (p_j - p_i)/L_e$, where $K_e$ is the dynamically adaptive conductivity,
$L_e$ is the length of the edge, and $p_i$ is the potential (e.g., blood pressure or
morphogen concentration) at unit $i$.
In plants, proteins embedded in the plasma membrane are responsible
for transporting auxin~\cite{Blakeslee2005,Kramer2006} with facilitated diffusion constants $K_e$.
In animals, blood flow through vessels can be approximated by Poiseuille's
law $K_e = k R_e^4$ with a constant $k$ and effective vessel radius $R_e$~\cite{Hacking1996,Hu2012}.

Let $\Delta: \mathcal N \rightarrow \mathcal E$ be the network's oriented incidence
matrix which maps the node vector space $\mathcal N$ to the edge vector space $\mathcal E$.
The matrix $\Delta$ acts as a discrete difference operator.
For each edge an arbitrary but fixed orientation is chosen (see Fig.~1 C in the main paper).
Then the components $\Delta_{e,i}$ read:
\begin{align}
	\Delta_{e,i} = \begin{cases}
		1, &\text{edge $e$ points towards node $i$} \\
        -1, &\text{edge $e$ points away from node $i$} \\
        0, &\text{edge $e$ is not connected to node $i$.}
	\end{cases}
\end{align}
The current vector $\mathbf F \in \mathcal E$ with entries $F_e$ can be derived from the potentials $\mathbf p \in \mathcal N$ using the formula
\begin{align}
	\mathbf F = K L^{-1}\; \Delta \mathbf p,
    \label{eq:flow}
\end{align}
The conductivities and lengths are summarized in the diagonal matrices $K$ and $L$.

The current balance at each node reads in vector form
\begin{align}
	\Delta^\top \mathbf F = \mathbf S,
    \label{eq:aux-balance}
\end{align}
where $\mathbf S$ is the source (or net current) term.
Eq.~\eqref{eq:aux-balance} is Kirchhoff's current law.
In plants, the source $\mathbf S$ describes the production rate of morphogen in each unit;
in animals, it represents the amount of blood perfusing one area unit.
Combining Eq.~\eqref{eq:flow} and Eq.~\eqref{eq:aux-balance}, we can solve for the steady state
currents and obtain
\begin{align}
	\mathbf{F} = K L^{-1} \Delta (\Delta^\top K L^{-1} \Delta)^\dagger \mathbf S,
    \label{eq:flow-sln}
\end{align}
where the dagger represents the Moore-Penrose pseudoinverse.
Equation \eqref{eq:flow-sln} can be used to compute the currents given all other
properties of the network.

\section{Nondimensionalization of the model}




Here, we explicitly derive the nondimensionalization of the dynamical equations
presented in the main paper.

The dimensionful dynamical equations used in the main paper are
\begin{align}
 \mathbf F &= K L^{-1}\Delta \left(\Delta^\top K L^{-1}\Delta\right)^\dagger
 	\mathbf S \\
 \frac{d K_e}{dt} &=a \left\langle F_e^2 \right\rangle^{\beta} - b K_e + c\exp(-rt),
 \label{eq:adapt-family}
\end{align}
where $\mathbf F$ is the flow state, a vector with elements $F_e$ corresponding to the flows through each edge $e$. The source term
$\mathbf S$ is the vector representing the net currents at each node. The parameters $a, b, c$ determine the adaptive dynamics,
$r$ is an inverse time scale of decay of the
background production term. $K_e$ is he conductivity of edge $e$. The angle brackets denote an average over the contributions from
all different fluctuating states.
Finally, we assume that there is a typical scale $\hat S$
for the source strengths.

We choose the following nondimensionalization:
\begin{align}
K = \frac{a}{b} \hat S^{2\beta} \tilde K, \quad &t = \frac{1}{b} \tilde t, \quad \mathbf F = \hat S \tilde{\mathbf F} \nonumber\\
L = \hat L \tilde L, \quad & \mathbf S = \hat S \tilde{\mathbf S}. \label{eq:typical-scales}
\end{align}
The definitions of the symbols follow the main paper:
$K$ is the vessel conductivity, $t$ is time, $\mathbf F$ is the current vector, $L$ is the vessel length,
and $\mathbf S$ is the source strength.
Quantities with a tilde are dimensionless and quantities with a hat are typical scales.
The model equations then reduce to the dimensionless system
\begin{align}
 \tilde{\mathbf F} &= \tilde K \tilde L^{-1}\Delta \left(\Delta^\top \tilde K \tilde L^{-1}\Delta\right)^\dagger
 	\tilde{\mathbf S} \\
 \frac{d \tilde K_e}{d\tilde t} &= \langle\tilde F_e^2\rangle^{\beta} - \tilde K_e + \kappa \exp(-\tilde t/\rho),
\end{align}
with the dimensionless control parameters $\kappa = (c/a) \hat S^{-2\beta}$,
$\rho=b/r$.

\section{Steady-state condition}
We simulate the network dynamics until a steady state condition is
reached. As a steady state measure, for each time step $i$ we compute the
dissipation
\begin{align*}
    E_i = \sum_{e} \frac{\langle F_e^2 \rangle}{K_e},
\end{align*}
where the sum runs over all edges with $K_e > 10^{-8}$ (the nonzero edges).
Then the simulation is ended once the relative change $|E_{i} - E_{i-1}|/|E_{i-1}| < 10^{-12}$,
which we take as the condition that steady state has been reached.

\section{Influence of the simulation parameters on distance from the Pareto front}
\begin{figure}
    \centering
    \includegraphics[width=.8\textwidth]{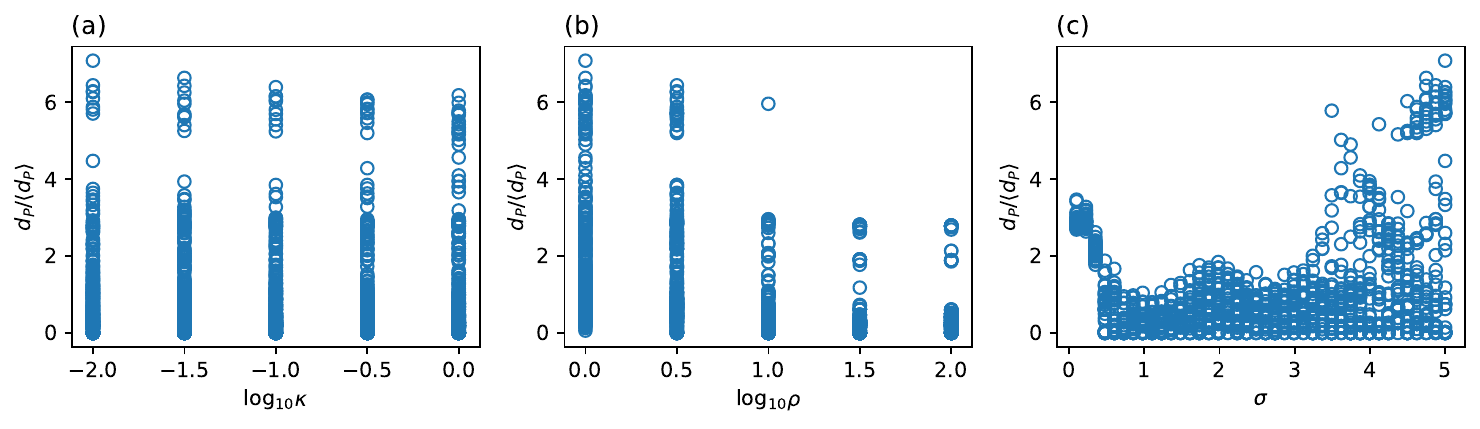}
    \caption{Distance from the Pareto front $d_P(x) = \min_{p\in P}
    \| x - p \|$ for the scaled data set analyzed in the main paper.
    Distances are normalized by the average $\langle d_P \rangle$ over
    all points not on the Pareto front.
    We plot the distance as a function of the three simulation parameters.
    (a) Increasing the growth strength $\kappa$ beyond $\kappa \approx 10$
    can drive networks closer to the Pareto front.
    (b) The time scale $\rho$ itself has little influence on distance from the
    Pareto front.
    (c) Fluctuation scales $0.5 \lesssim \sigma \lesssim 3$ lead to networks
    much closer to the Pareto front than other values.}
    \label{fig:pareto-distance}
\end{figure}

In Fig.~\ref{fig:pareto-distance} we show the influence of the simulation
parameters on the average distance from the Pareto front for the
data set analyzed in the main paper.
As a function of $\rho$, the distance is essentially random.
As a function of $\kappa$, it is random for low values of $\kappa$,
but networks are driven closer to the Pareto front for larger $\kappa$.
As a function of $\sigma$, networks are close to the Pareto front
for a range of medium values $0.5 \lesssim \sigma \lesssim 3$.

\section{Algorithm for computing the Pareto front}
Given a set of $n$ observations of $m$ objectives $Y = \{ \mathbf{y}_n \}_{n}$
to be minimized, where 
$\mathbf{y} = (y_1,\dots, y_m)$, we can introduce a partial ordering
by defining
$\mathbf{y} \prec \mathbf{y}'$ if $y_i \leq y_i'$ for all $i$, and $y_j < y_j'$
for at least one $j$. We then say that $\mathbf{y}$ dominates $\mathbf{y}'$.
The \emph{Pareto front} is then the set
\begin{align}
    P = \{ \mathbf{y} \in Y \mid D(\mathbf{y}) \text{ is empty} \},
\end{align}
where the set 
\begin{align}
    D(\mathbf{y}) = \{ \mathbf{y}' \in Y \mid \mathbf{y}' \neq \mathbf{y} \text{ and }
    \mathbf{y}' \prec \mathbf{y} \}
\end{align}
is the set of all points that dominate $\mathbf{y}$.
Thus, the Pareto front is the set of all points that are not dominated 
by any other points.
In order to find the Pareto front, we follow Ref.~[39] from the main paper
and implement their Algorithm 2 (Simple Cull).
For reference, we reproduce pseudocode in Algorithm~\ref{alg:pareto}.

\begin{figure}
\begin{algorithm}[H]
  \caption{Simple Cull algorithm for finding the 
  Pareto front $P$, from Ref.~[39].}
  \label{alg:pareto}
   \begin{algorithmic}[1]
   \State $P := \{\}$ 
   
   \While{$Y \neq \{\}$}
    \State $y := $ \texttt{RemoveElementFrom}$(Y)$
    \State \texttt{dominated} $:=$ \texttt{False}
    
    \For{each $d \in P$}
        \If{$c \prec d$}
            \State $P := P \setminus \{d\}$
        \Else
            \State \texttt{dominated} $:=$ \texttt{True},
            \textbf{break}.
        \EndIf
    \EndFor
    
    \If {not \texttt{dominated}}
        \State $P := P \cup \{c\}$
    \EndIf
   \EndWhile
   \end{algorithmic}
\end{algorithm}
\end{figure}

\section{Principal Component Analysis}

Given an $n\times m$ matrix of data $\mathbf{Y}$ with $n$
observations of $m$ objectives, where each column has
mean $0$ and variance $1$, Principal Component Analysis
computes the eigen-decomposition of $\mathbf{Y}^\top \mathbf{Y}$
(proportional to the covariance matrix of the data).
The eigenvalues are then ordered from largest to smallest,
and are proportional to the fraction of
the total variance encoded in the component of the data
in the direction of the corresponding eigenvector.
If $\mathbf{v}_i$ is the $i$'th PCA eigenvector, then
the corresponding $i$'th PCA coordinate (PCA $i$) of a point
$\mathbf{y}$ (one row of $\mathbf{Y}$) is
the inner product $\mathbf{v}_i^\top \mathbf{y}$.

In the main paper, the data matrix $\mathbf{Y}$ consists
of the points on the Pareto front found using
Algorithm~\ref{alg:pareto}.
The PCA coordinates are then still well-defined for \emph{any} point
(not just Pareto points) and are computed using 
$\mathbf{v}_i^\top \mathbf{y}$.

\section{Results for other boundary conditions and lattices}
In this section we show computational results for other lattices,
boundary conditions, and fluctuation functions.
In all cases, we use the same parameter values as in the main paper.

The collectively produced fluctuations of the sources are generated by a function of the form:
\begin{align}
\frac{\mathbf{S}^{(i)}_j}{\hat S} = \delta_{j0} - (1 - \delta_{j0})\, f\left(\frac{\|\mathbf x_j - \mathbf x_i\|}{\sigma} \right),
\label{eq:fluct-new}
\end{align}
where $\mathbf x_i$ is the position of node $i$, $\sigma$ is the scale over which the source strength
varies, and
$\sum_j \mathbf S^{(i)}_j = 0$. The function $f(x)$ determines the type of fluctuations. The total in- and outflow is $\hat S$.

\subsection{Gaussian sources}
Here, we show results for Gaussian collective sources where the inlet is
located at the left side of the network, in addition to the case considered in
the main paper
Gaussian sources are given by
\begin{align}
f(x) \sim e^{-\frac{x^2}{2}},
\end{align}
where the constant of proportionality
is computed from the condition $\sum_j S_j^{(i)} = 0$.

\begin{figure}
    \centering
    \includegraphics[width=0.6\textwidth]{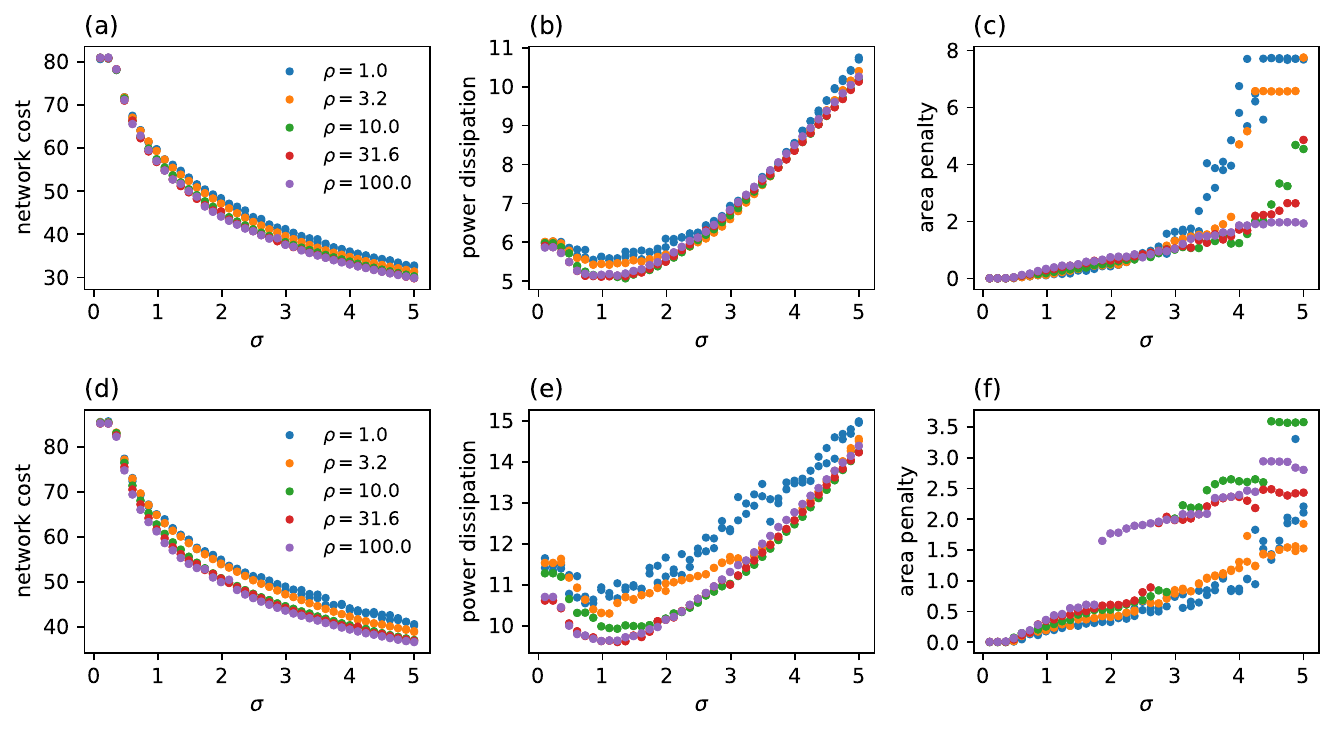}
    \caption{Network measures for Gaussian sources at fixed $\kappa=1.0$.
    (a--c) A single inlet at the center of the network.
    (d--f) A single inlet at the left side of the network.}
    \label{fig:measures_gaussian}
\end{figure}

\begin{figure}
    \centering
    \includegraphics[width=\textwidth]{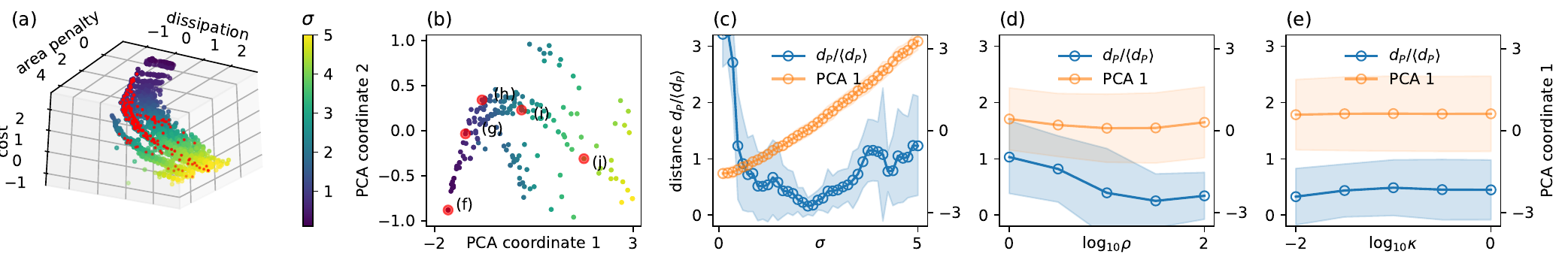}
    \caption{Equivalent of Fig.~3 from the main paper for the case of
    Gaussian sources with an inlet at the left boundary of the network.
    The Pareto front shows distinctive branches which individually
    resemble one-dimensional curves.
    The other results are qualitatively similar to those for a centered inlet
    from the main paper.
    The marked networks in panel (b) correspond to the networks
    (f--j) from Fig.~2 in the main paper.}
    \label{fig:fig3_leaf}
\end{figure}

The network phenotypes show the same qualitative behavior as for the case of the main paper,
where the source was at the center.
The one exception is the percolation penalty (Fig.~\ref{fig:measures_gaussian}~(f)), which shows two branches.
This is because as $\sigma$ increases and the number of loops decreases, it becomes
more likely that the source is only connected by a single edge
to the rest of the network. In contrast, with the source in the center,
this is unlikely to happen.
This branching can also be seen in the phenotypic space and in the
corresponding Pareto front (Fig.~\ref{fig:fig3_leaf}).
Otherwise, the case of a single inlet at the left qualitatively agrees
with the one from the main paper.

\begin{figure}
    \centering
    \includegraphics[width=\textwidth]{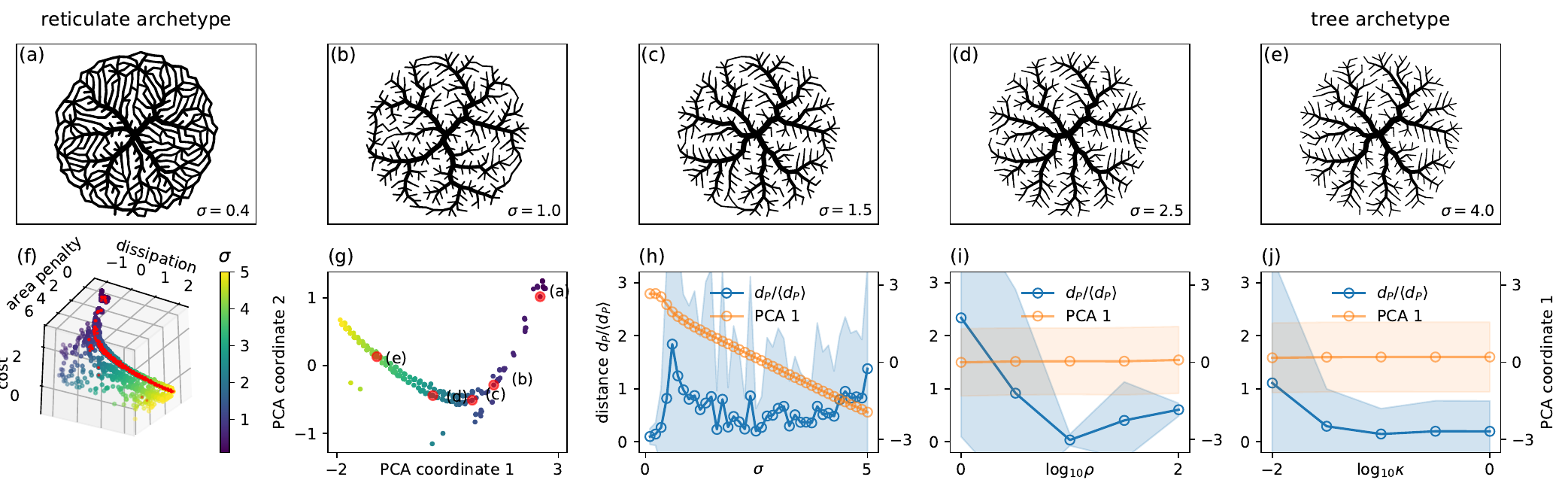}
    \caption{Results for $\beta=0.77$ with Gaussian sources and an inlet
    at the center of the network.
    (a--e) Network phenotypes along the Pareto front.
    (f) Task space with Pareto front in red.
    (g) PCA embedding of the Pareto front shows approximately 1-dimensional curve
    Labels (a--e) correspond to the networks shown in corresponding panels.
    (h--j) Equivalents to Fig.~3 (d--f) in the main paper show that
    $\sigma$ parametrizes the Pareto front while $\kappa$ and $\rho$ do not.}
    \label{fig:beta_077}
\end{figure}

\begin{figure}
    \centering
    \includegraphics[width=\textwidth]{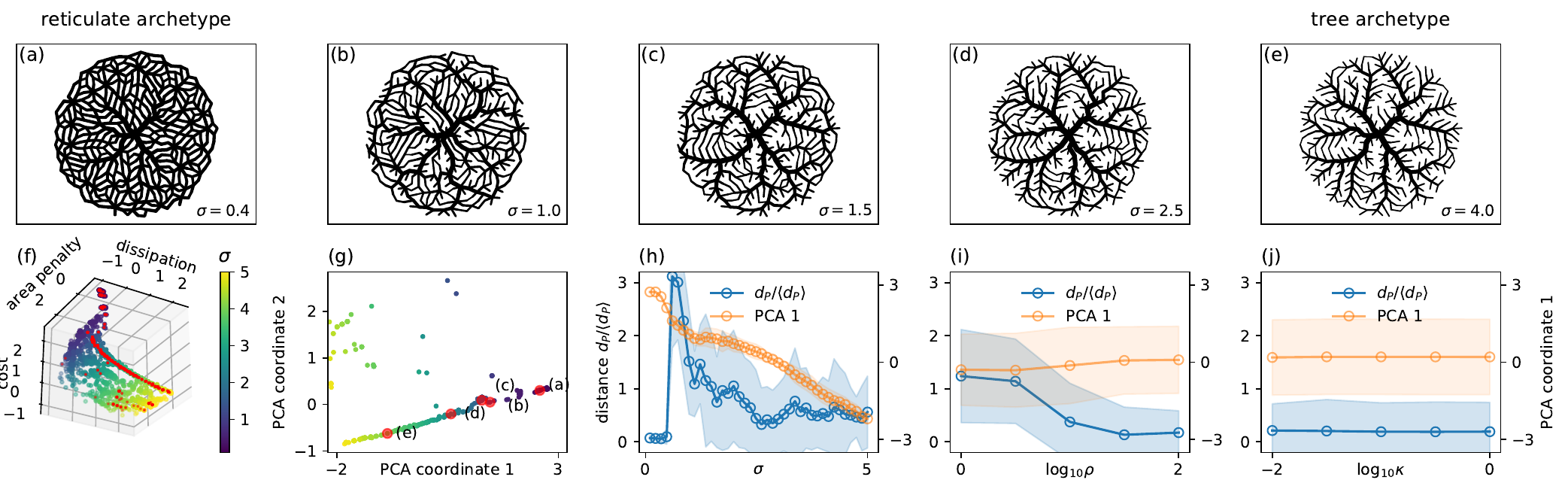}
    \caption{Results for $\beta=0.63$ with Gaussian sources and an inlet
    at the center of the network.
    (a--e) Network phenotypes along the Pareto front.
    (f) Task space with Pareto front in red.
    (g) PCA embedding of the Pareto front shows approximately 1-dimensional curve
    Labels (a--e) correspond to the networks shown in corresponding panels.
    (h--j) Equivalents to Fig.~3 (d--f) in the main paper show that
    $\sigma$ parametrizes the Pareto front while $\kappa$ and $\rho$ do not.}
    \label{fig:beta_063}
\end{figure}

Furthermore, we investigated different values of the nonlinearity parameter
$\beta$ in the biologically relevant regime $\beta > 1/2$ (Smaller $\beta$ lead to
fully reticulate networks in all cases). 
Specifically, we looked that the case of center inlets.
The results for $\beta=0.77$ are shown in Fig.~\ref{fig:beta_077},
and the results for $\beta=0.63$ are shown in Fig.~\ref{fig:beta_063}.
Qualitatively, we obtain the same results as for $\beta=2/3$, the case considered
in the main paper.

In all cases (main paper and supplement), we scanned a parameter range
of 5 logarithmically distributed points between $\rho = 1$ and $\rho = 100$,
5 logarithmically distributed points between $\kappa = 0.01$ and $\kappa = 1$, and
40 linearly distributed points between $\sigma = 0.1$ and $\sigma = 1$.
For each combination, we took 2 samples with random initial conditions.

\subsection{Exponential sources}
\begin{figure}
    \centering
    \includegraphics[width=.98\textwidth]{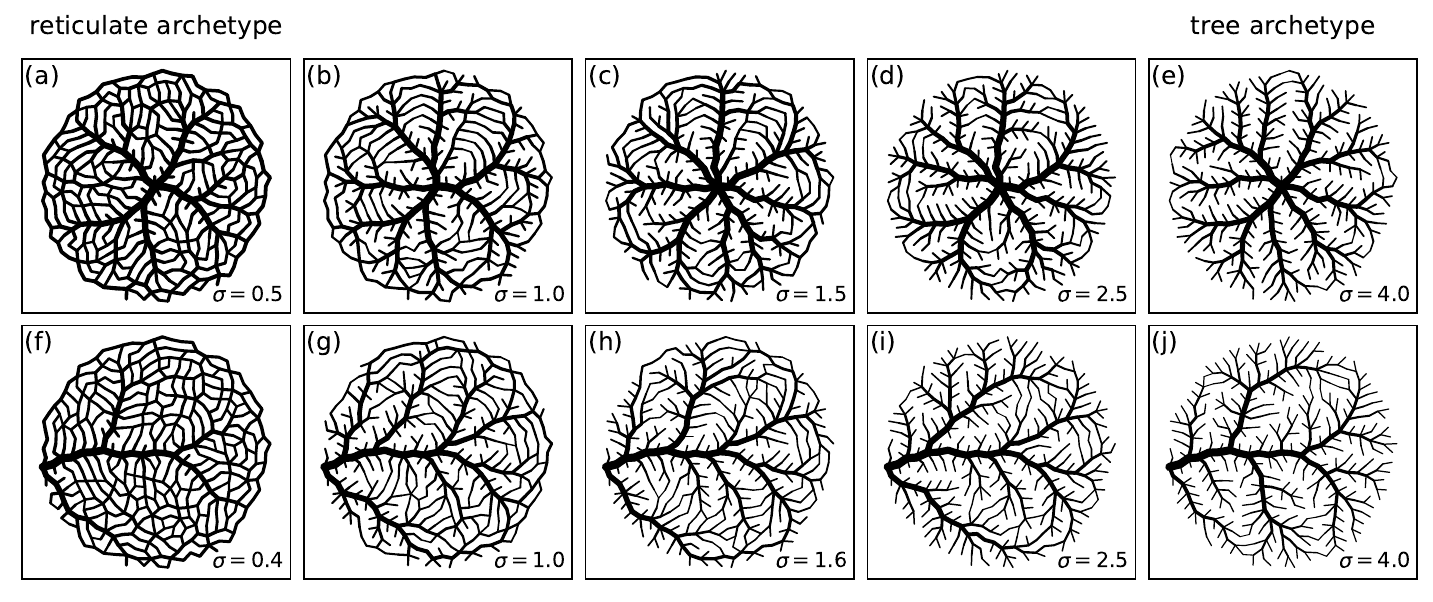}
    \caption{The equivalent of Fig.~2 from the main paper for exponential
    sources. Networks on the Pareto front behave qualitatively analogously
    to Gaussian sources.}
    \label{fig:fig2_exponential}
\end{figure}

In this subsection we show results for exponential collective sources,
\begin{align}
f(x) \sim e^{-x}.
\end{align}
We find no qualitative difference between exponential and Gaussian distributed sources
in the phenotypes as a function of $\sigma$ or the phenotypic space.
Fig.~\ref{fig:fig2_exponential} shows the equivalent of Fig.~2 from the main
paper for exponential sources.

\begin{figure}
    \centering
    \includegraphics[width=\textwidth]{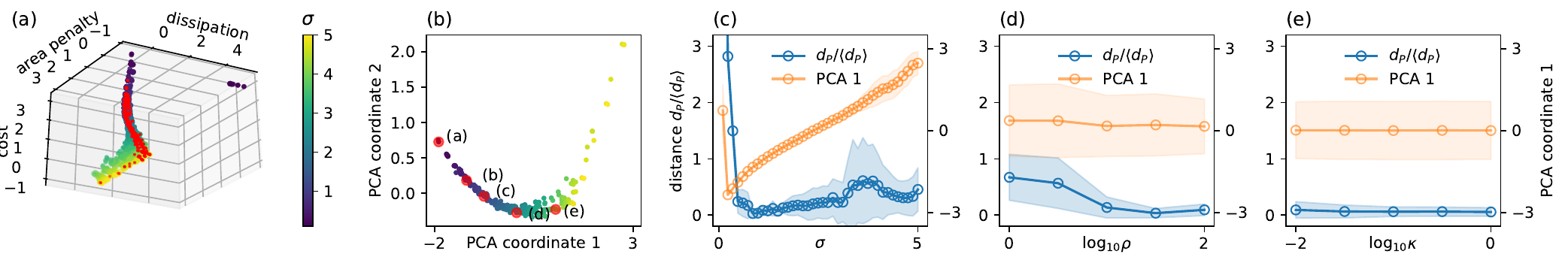} \\
    \caption{Equivalent of Fig.~3 from the main paper for the case of
    exponential sources with an inlet at the center of the network.
    Results are qualitatively similar to those
    from the main paper.
    The marked networks in panel (b) correspond to the networks
    (a--e) from Fig.~\ref{fig:fig2_exponential}.}
    \label{fig:fig3_exponential_animal}
\end{figure}

\begin{figure}
    \centering
    \includegraphics[width=\textwidth]{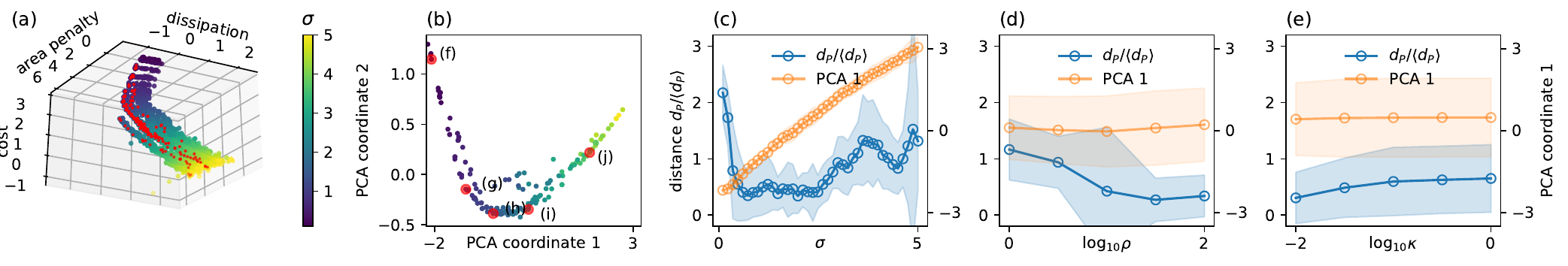} \\
    \caption{Equivalent of Fig.~3 from the main paper for the case of
    exponential sources with an inlet at the boundary the network.
    Results are qualitatively similar to those
    from the main paper.
    The marked networks in panel (b) correspond to the networks
    (f--j) from Fig.~\ref{fig:fig2_exponential}.}
    \label{fig:fig3_exponential_leaf}
\end{figure}

\begin{figure}
    \centering
    \includegraphics[width=.6\textwidth]{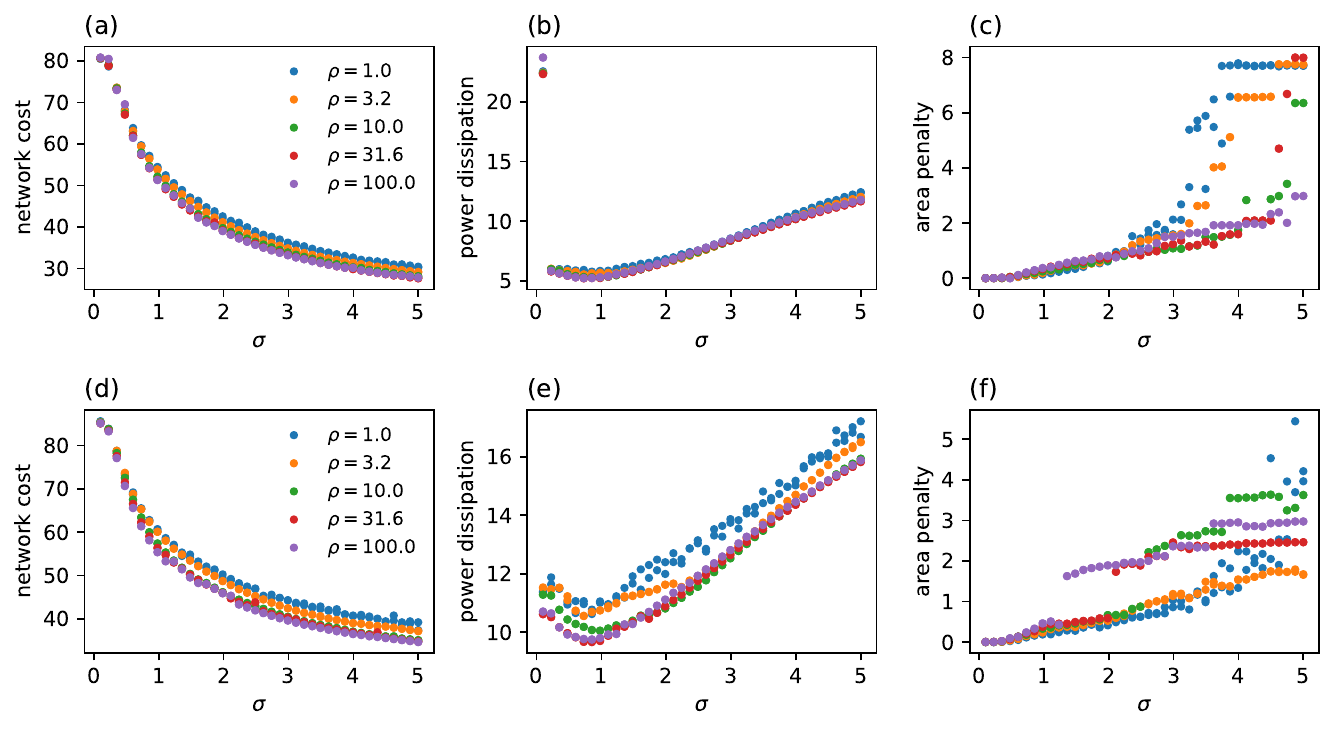}
    \caption{Network measures for exponential sources at fixed $\kappa=1.0$.
    (a--c) A single inlet at the center of the network.
    (d--f) A single inlet at the left side of the network.}
    \label{fig:measures_exponential}
\end{figure}

Figs.~\ref{fig:fig3_exponential_animal} and \ref{fig:fig3_exponential_leaf}
correspond to Fig.~3 from the main paper for inlets at the
center and on the boundary of the network, respectively. Again we
find qualitatively similar results to the case investigated in the main paper.
For completeness, we show the dependence of the network measures
on correlation length at fixed $\kappa$ in Fig.~\ref{fig:measures_exponential}.

In all cases, we scanned a parameter range
of 5 logarithmically distributed points between $\rho = 1$ and $\rho = 100$,
5 logarithmically distributed points between $\kappa = 0.01$ and $\kappa = 1$, and
40 linearly distributed points between $\sigma = 0.1$ and $\sigma = 1$.
For each combination, we took 2 samples with random initial conditions.

\subsection{Random sources}
In this subsection we show results from random fluctuations.
We consider fluctuating states
\begin{align}
S^{(i)}_j = \begin{cases}
-k & \text{with probability } p \\
0 & \text{with probability } 1 - p
\end{cases}
\end{align}
for $j = 1,\dots, N-1$. The source is normalized to $(S_i)_0 = \sum_{j>0} (S_i)_j = 1$.
This normalization sets the value of of the constant $k$.
In the case of these random fluctuations, we consider fluctuation averages over
100 different arrangements of the sources that all satisfy the above probabilistic
condition.
For these types of fluctuations, the resulting network often does not connect each node
in the original lattice to the source node. Thus, our metrics such as the percolation
penalty, network cost, and uniform energy cannot be directly compared anymore.
Yet, inspecting the simulation results for various values of $p$ shows that
this model does not reproduce strongly hierarchically ordered networks,
see Fig.~\ref{fig:random-networks-leaf} for the case of leaf-like boundary conditions
as an example.
%

\begin{figure}
\includegraphics[width=.6\columnwidth]{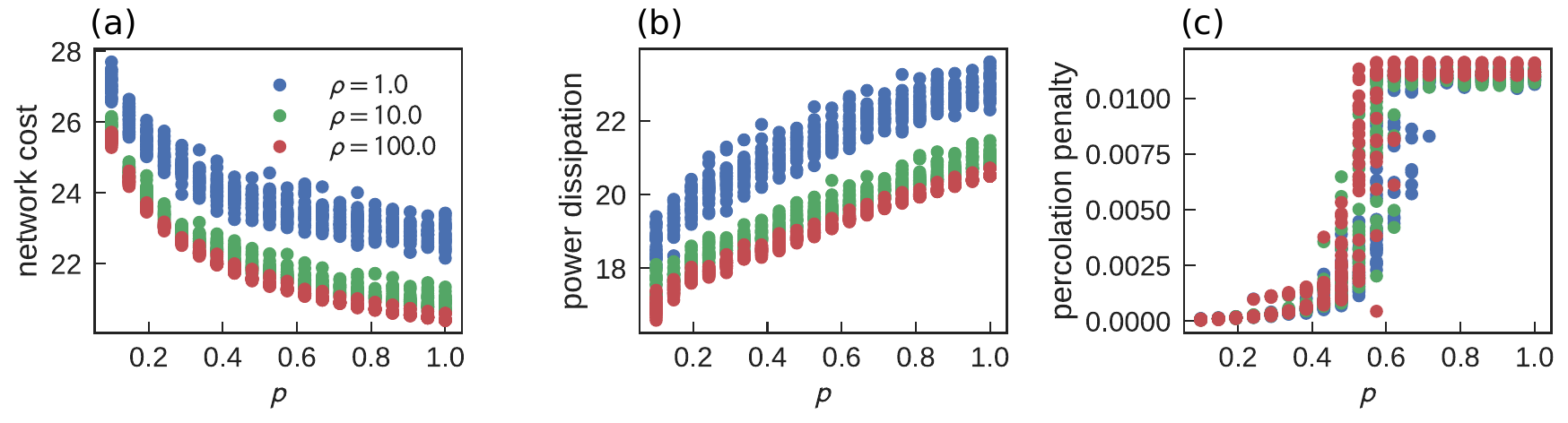}
\caption{Dependence of network phenotypes on probability for
randomly distributed fluctuations with one source at the
left boundary for fixed $\kappa=1.0$. Two distinct phases can be discerned in the percolation penalty
with a sharp transition between them.
For $p>0.5$, no more loops are produced in the networks such that the percolation
penalty remains at a large value. For $p<0.5$, the networks are well-connected.\label{FigS9}
\label{fig:random-dependence}}
\end{figure}

\begin{figure}
\includegraphics[width=.7\columnwidth]{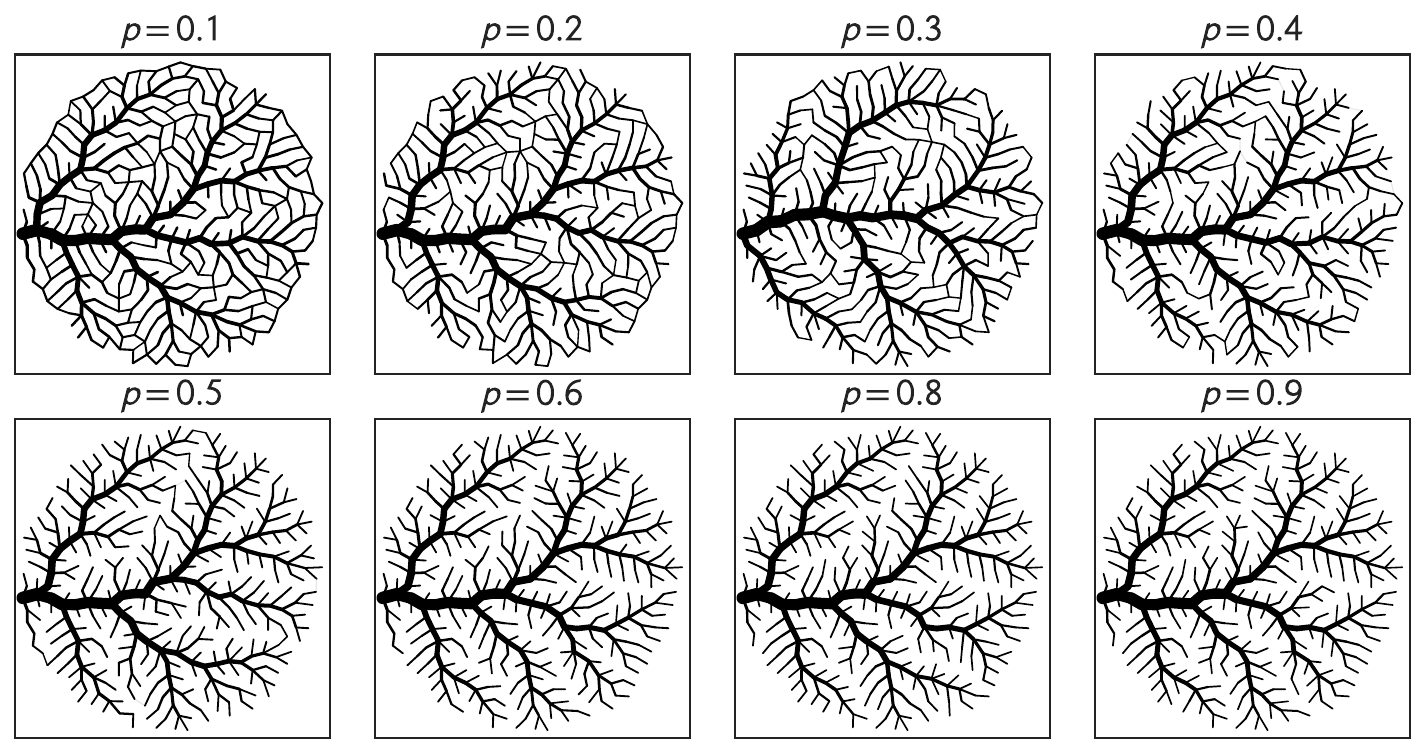}
\caption{\label{fig:random-networks-leaf}Random networks for a source at the left boundary and various
probabilities $p$ of any node being a source. For $p>0.5$, the networks are
trees.
We used parameter values $\kappa=1, \rho=99$.}
\end{figure}

Comparing also with Fig.~\ref{fig:random-dependence}, we observe a sharprhos transition
between a well-connected network for $p<0.5$ and a minimally connected
topological tree for $p>0.5$. The case $p=0.5$ was considered in detail in Ref.~\cite{Corson2010}.

In all cases, we scanned a parameter range
of 5 logarithmically distributed points between $\rho = 1$ and $\rho = 100$,
5 logarithmically distributed points between $\kappa = 0.01$ and $\kappa = 1$, and
40 linearly distributed points between $p = 0.1$ end $p = 1$.
For each combination, we took 2 samples with random initial conditions.

\subsection{Uniform + moving point sources}
Here we consider a source term that is composed of a contribution from
uniform sinks on the network, and one from random sinks.
Specifically, we consider
\begin{align}
    S_j^{(i)} = \delta_{j0} - \alpha\,(1 - \delta_{j0}) (1 - e + e\,\delta_{ij}).
\end{align}
The parameter $e$ controls the relative importance of both terms,
and the normalization factor $\alpha$ is chosen to enforce $\sum_j S_j^{(i)} = 0$.
If $e=0$, there are uniform sinks and we expect a hierarchical tree network.
If $e=1$, the sinks are fully random, and we expect a fully reticulate
network analogous to the $\sigma \to 0$ case for a distance-dependent fluctuation
function.
Numerically, we observe that networks remain trees until approximately
$e\approx 0.96$. Hence, we explore the parameter range $0.96 \leq e\leq 1.0$
in detail.

\begin{figure}
    \centering
    \includegraphics[width=.98\textwidth]{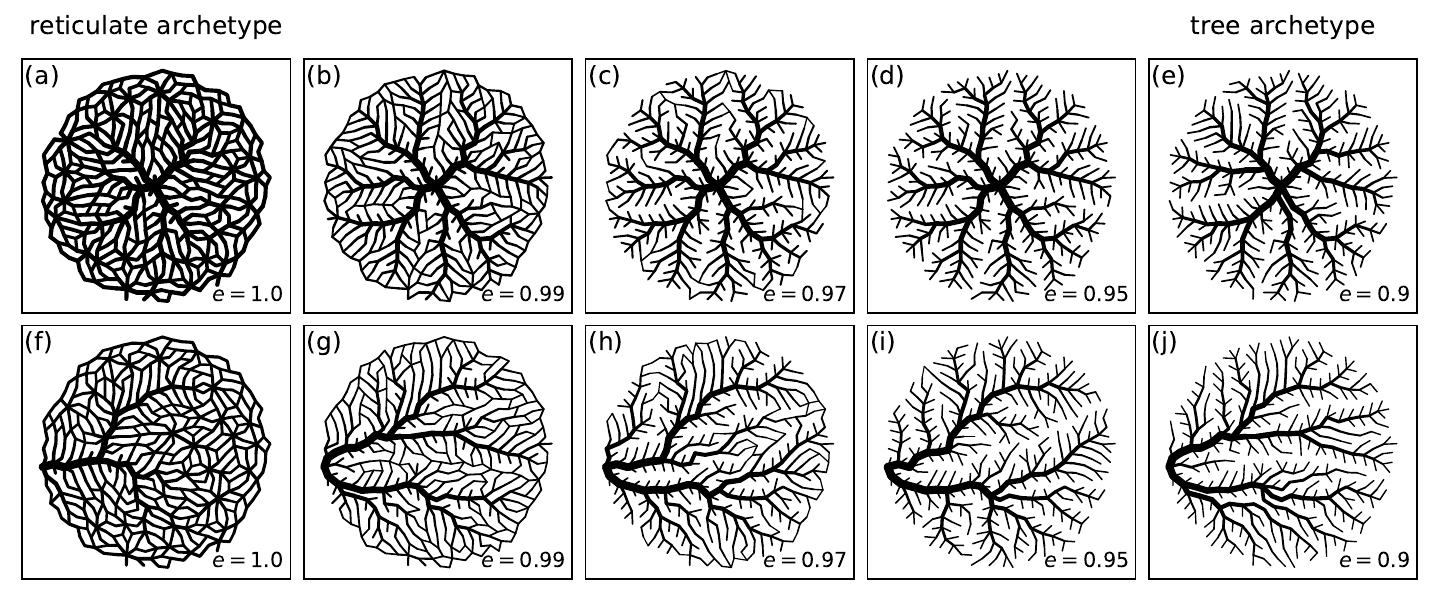}
    \caption{The equivalent of Fig.~2 from the main paper for the
    uniform sources + moving sink model. Networks on the Pareto front behave qualitatively analogously
    to Gaussian sources, as discussed in the main paper, and
    the parameter $e$ controls reticulation.}
    \label{fig:fig2_unif}
\end{figure}

\begin{figure}
    \centering
    \includegraphics[width=.98\textwidth]{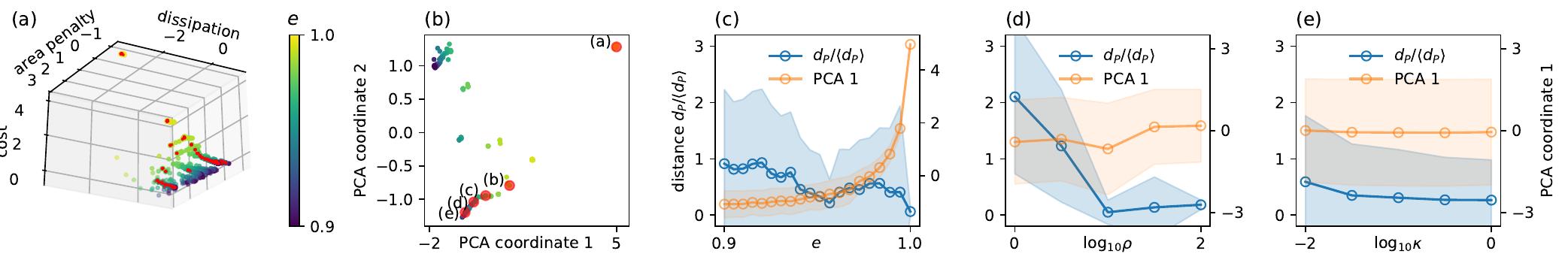}
    \caption{Equivalent of Fig.~3 from the main paper for the case of
    uniform + moving point sources with an inlet at the center of the network.
    Results are qualitatively similar to those
    from the main paper in terms of parametrizing the 
    Pareto front, but the shape of the Pareto front
    shows distinct clusters, and is not 1-dimensional.
    The marked networks in panel (b) correspond to the networks
    (a--e) from Fig.~\ref{fig:fig2_unif}.}
    \label{fig:fig3_unif_center}
\end{figure}

\begin{figure}
    \centering
    \includegraphics[width=.98\textwidth]{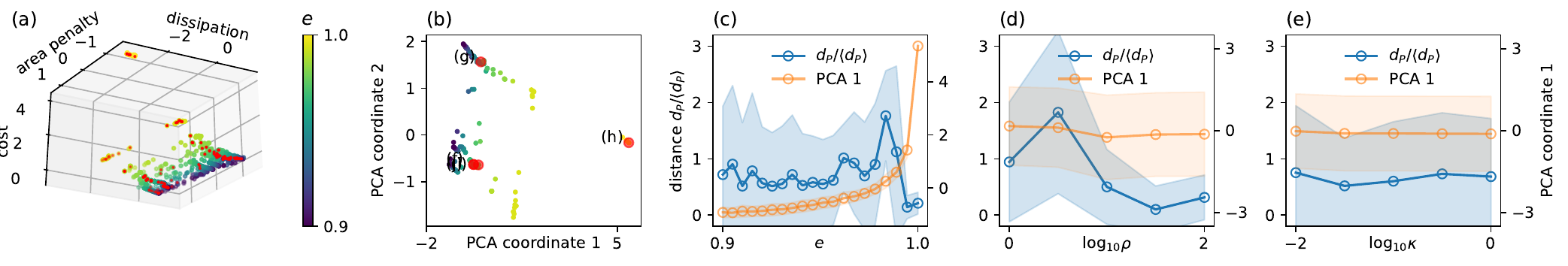}
    \caption{Equivalent of Fig.~3 from the main paper for the case of
    uniform + moving point sources with an inlet at the left of the network.
    Results are qualitatively similar to those
    from the main paper in terms of parametrizing the 
    Pareto front, but the shape of the Pareto front
    shows distinct clusters, and is not 1-dimensional.
    The marked networks in panel (b) correspond to the networks
    (f--j) from Fig.~\ref{fig:fig2_unif}.}
    \label{fig:fig3_unif_left}
\end{figure}

Fig.~\ref{fig:fig2_unif} corresponds to Fig. 2 from the main paper, and
Figs.~\ref{fig:fig3_unif_center} and \ref{fig:fig3_unif_left}
correspond to Fig.~3 from the main paper for inlets at the
center or at the left. Again we
find qualitatively similar results to the case investigated in the main paper when it comes to the ability of the parameter
$e$ to interpolate along the Pareto front.
However, the Pareto front is more fragmented and shows distinct
clusters as opposed to the clear approximately 1-dimensional
geometry found for Gaussian or exponential sources.

In all cases, we scanned a parameter range
of 5 logarithmically distributed points between $\rho = 1$ and $\rho = 100$,
5 logarithmically distributed points between $\kappa = 0.01$ and $\kappa = 1$, and
20 logarithmically distributed points between $e = 0.9$ end $e = 1$.
For each combination, we took 2 samples with random initial conditions.

\section{Fluctuation average and eigendecomposition of the source covariances}
As shown the the preceding section, the vector of flows can be written
as a linear map acting on the vector of sources,
\begin{align}
 \mathbf F &= K L^{-1}\Delta \left(\Delta^\top K L^{-1}\Delta\right)^\dagger
 	\mathbf S \\
 	&= A \, \mathbf S.
\end{align}
The fluctuation average over several sources can therefore be expressed as
\begin{align}
    \langle F_e^2 \rangle &= \frac{1}{N} \sum_i (\mathbf e^\top A \mathbf S^{(i)})^2 \\
    &=\mathbf e^\top A \underbrace{\frac{1}{N} \left( \sum_i \mathbf S^{(i)} \mathbf (\mathbf S^{(i)})^\top \right)}_{= C} A^\top \mathbf{e},
\end{align}
where $\mathbf e$ is the unit vector corresponding to edge $e$.
The matrix $C$ is precisely the matrix of (uncentered) covariances between the source states.
Using the eigendecomposition $C = \sum_j \rho_j \mathbf r_j \mathbf r_j^\top$,
we find
\begin{align}
    \langle F_e^2 \rangle
    &=\mathbf e^\top A \sum_j \rho_j \mathbf r_j \mathbf r_j^\top  A^\top \mathbf{e} \\
    &= \sum_j \rho_j (\mathbf e^\top A \mathbf{r}_j)^2,
\end{align}
which corresponds to Eq.~(5) in the main paper with $R^{(j)}_e = \mathbf e^\top A \mathbf{r}_j$.

\begin{figure}
    \centering
    \includegraphics[width=\textwidth]{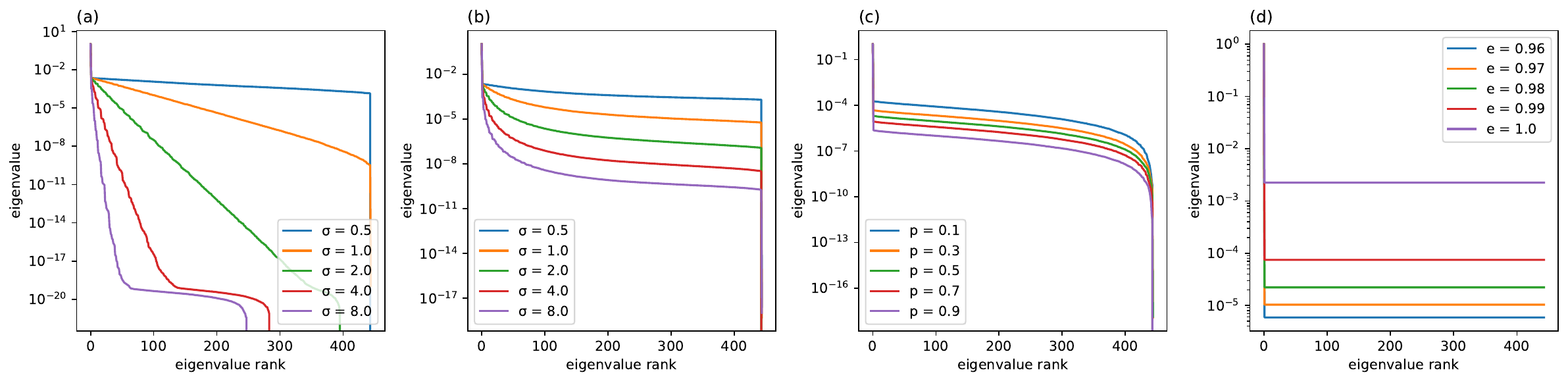}
    \caption{The eigenvalues of the correlation
    matrix $C$ for the network topology considered here and in the main paper,
    with an inlet at the center of the network.
    (a) Gaussian sources (b) exponential sources (c) random sources 
    (d) uniform + moving point sources.
    All correlation matrices are characterized by a single, dominant
    eigenvalue whose eigenvector approximately corresponds to
    a single inlet uniform with uniform sinks.
    When the parameters are chosen such that the other eigenvalues are
    small, only this dominant eigenvalue is relevant, and leads to
    a hierarchical tree network topology.}
    \label{fig:correlation}
\end{figure}

The eigenvalues of the correlation matrix $C$
generally decay rapidly for large
correlation lengths $\sigma$, as shown in
Fig.~\ref{fig:correlation}.

\section{Covariance matrix for uncorrelated fluctuations}
As long as the fluctuation function satisfies $f(0) = 1$ and
$f(\infty) = 0$, the limit
$\sigma\to 0$ will lead to uncorrelated fluctuations in the
statistical sense.
The sources become
\begin{align*}
    S_j^{(i)} = \delta_{j0} + (1 - \delta_{j0})\delta_{ji},
\end{align*}
which corresponds to a single sink randomly placed at node $i$.
We now proceed to calculate the covariance matrix between
different node sinks.

Statistically, each of the $N$ sink nodes is active with probability $1/N$ and strength 1.
Thus,
\begin{align}
    \langle S_j \rangle = \frac{1}{N}
\end{align}
for $j>0$. We can further calculate
\begin{align}
    \langle S_i S_j \rangle = \frac{1}{N} \frac{1}{N}
\end{align}
for $i\neq j$ and $i,j>0$ because the sources are independent. Because $\langle S_i^2\rangle = \frac{1}{N}$, we find the general expression
\begin{align}
    \langle S_i S_j \rangle = \frac{1}{N^2} + \left( \frac{1}{N}  - \frac{1}{N^2} \right)
    \delta_{ij},
\end{align}
such that finally, the statistical covariance matrix becomes
\begin{align}
\langle S_i S_j \rangle - \langle S_i\rangle \langle S_j\rangle = \left( \frac{1}{N}  - \frac{1}{N^2} \right)
    \delta_{ij}, 
\end{align}    
which is proportional to the unit matrix and therefore
corresponds to uncorrelated fluctuations.
    
\begin{thebibliography}{44}%
\makeatletter
\providecommand \@ifxundefined [1]{%
 \@ifx{#1\undefined}
}%
\providecommand \@ifnum [1]{%
 \ifnum #1\expandafter \@firstoftwo
 \else \expandafter \@secondoftwo
 \fi
}%
\providecommand \@ifx [1]{%
 \ifx #1\expandafter \@firstoftwo
 \else \expandafter \@secondoftwo
 \fi
}%
\providecommand \natexlab [1]{#1}%
\providecommand \enquote  [1]{``#1''}%
\providecommand \bibnamefont  [1]{#1}%
\providecommand \bibfnamefont [1]{#1}%
\providecommand \citenamefont [1]{#1}%
\providecommand \href@noop [0]{\@secondoftwo}%
\providecommand \href [0]{\begingroup \@sanitize@url \@href}%
\providecommand \@href[1]{\@@startlink{#1}\@@href}%
\providecommand \@@href[1]{\endgroup#1\@@endlink}%
\providecommand \@sanitize@url [0]{\catcode `\\12\catcode `\$12\catcode
  `\&12\catcode `\#12\catcode `\^12\catcode `\_12\catcode `\%12\relax}%
\providecommand \@@startlink[1]{}%
\providecommand \@@endlink[0]{}%
\providecommand \url  [0]{\begingroup\@sanitize@url \@url }%
\providecommand \@url [1]{\endgroup\@href {#1}{\urlprefix }}%
\providecommand \urlprefix  [0]{URL }%
\providecommand \Eprint [0]{\href }%
\providecommand \doibase [0]{http://dx.doi.org/}%
\providecommand \selectlanguage [0]{\@gobble}%
\providecommand \bibinfo  [0]{\@secondoftwo}%
\providecommand \bibfield  [0]{\@secondoftwo}%
\providecommand \translation [1]{[#1]}%
\providecommand \BibitemOpen [0]{}%
\providecommand \bibitemStop [0]{}%
\providecommand \bibitemNoStop [0]{.\EOS\space}%
\providecommand \EOS [0]{\spacefactor3000\relax}%
\providecommand \BibitemShut  [1]{\csname bibitem#1\endcsname}%
\let\auto@bib@innerbib\@empty
\bibitem [{\citenamefont {Ronellenfitsch}\ \emph {et~al.}(2015)\citenamefont
  {Ronellenfitsch}, \citenamefont {Lasser}, \citenamefont {Daly},\ and\
  \citenamefont {Katifori}}]{Ronellenfitsch2015b}%
  \BibitemOpen
  \bibfield  {author} {\bibinfo {author} {\bibfnamefont {Henrik}\ \bibnamefont
  {Ronellenfitsch}}, \bibinfo {author} {\bibfnamefont {Jana}\ \bibnamefont
  {Lasser}}, \bibinfo {author} {\bibfnamefont {Douglas~C.}\ \bibnamefont
  {Daly}}, \ and\ \bibinfo {author} {\bibfnamefont {Eleni}\ \bibnamefont
  {Katifori}},\ }\bibfield  {title} {\enquote {\bibinfo {title} {{Topological
  Phenotypes Constitute a New Dimension in the Phenotypic Space of Leaf
  Venation Networks}},}\ }\href {\doibase 10.1371/journal.pcbi.1004680}
  {\bibfield  {journal} {\bibinfo  {journal} {PLOS Computational Biology}\
  }\textbf {\bibinfo {volume} {11}},\ \bibinfo {pages} {e1004680} (\bibinfo
  {year} {2015})},\ \Eprint {http://arxiv.org/abs/1507.04487}
  {arXiv:1507.04487} \BibitemShut {NoStop}%
\bibitem [{\citenamefont {Miettinen}(1999)}]{Miettinen1999}%
  \BibitemOpen
  \bibfield  {author} {\bibinfo {author} {\bibfnamefont {Kaisa}\ \bibnamefont
  {Miettinen}},\ }\href {https://books.google.com/books?id=ha\_zLdNtXSMC}
  {\emph {\bibinfo {title} {Nonlinear Multiobjective Optimization}}},\
  International Series in Operations Research \& Management Science\ (\bibinfo
  {publisher} {Springer US},\ \bibinfo {year} {1999})\BibitemShut {NoStop}%
\bibitem [{\citenamefont {Shoval}\ \emph {et~al.}(2012)\citenamefont {Shoval},
  \citenamefont {Sheftel}, \citenamefont {Shinar}, \citenamefont {Hart},
  \citenamefont {Ramote}, \citenamefont {Mayo}, \citenamefont {Dekel},
  \citenamefont {Kavanagh},\ and\ \citenamefont {Alon}}]{Shoval2012}%
  \BibitemOpen
  \bibfield  {author} {\bibinfo {author} {\bibfnamefont {O.}~\bibnamefont
  {Shoval}}, \bibinfo {author} {\bibfnamefont {H.}~\bibnamefont {Sheftel}},
  \bibinfo {author} {\bibfnamefont {G.}~\bibnamefont {Shinar}}, \bibinfo
  {author} {\bibfnamefont {Y.}~\bibnamefont {Hart}}, \bibinfo {author}
  {\bibfnamefont {O.}~\bibnamefont {Ramote}}, \bibinfo {author} {\bibfnamefont
  {A.}~\bibnamefont {Mayo}}, \bibinfo {author} {\bibfnamefont {E.}~\bibnamefont
  {Dekel}}, \bibinfo {author} {\bibfnamefont {K.}~\bibnamefont {Kavanagh}}, \
  and\ \bibinfo {author} {\bibfnamefont {U.}~\bibnamefont {Alon}},\ }\bibfield
  {title} {\enquote {\bibinfo {title} {{Evolutionary Trade-Offs, Pareto
  Optimality, and the Geometry of Phenotype Space}},}\ }\href {\doibase
  10.1126/science.1217405} {\bibfield  {journal} {\bibinfo  {journal}
  {Science}\ }\textbf {\bibinfo {volume} {336}},\ \bibinfo {pages} {1157--1160}
  (\bibinfo {year} {2012})},\ \Eprint {http://arxiv.org/abs/9605103}
  {arXiv:9605103 [cs]} \BibitemShut {NoStop}%
\bibitem [{\citenamefont {Givnish}\ \emph {et~al.}(2005)\citenamefont
  {Givnish}, \citenamefont {Pires}, \citenamefont {Graham}, \citenamefont
  {McPherson}, \citenamefont {Prince}, \citenamefont {Patterson}, \citenamefont
  {Rai}, \citenamefont {Roalson}, \citenamefont {Evans}, \citenamefont {Hahn},
  \citenamefont {Millam}, \citenamefont {Meerow}, \citenamefont {Molvray},
  \citenamefont {Kores}, \citenamefont {O'Brien}, \citenamefont {Hall},
  \citenamefont {Kress},\ and\ \citenamefont {Sytsma}}]{Givnish2005}%
  \BibitemOpen
  \bibfield  {author} {\bibinfo {author} {\bibfnamefont {T.~J}\ \bibnamefont
  {Givnish}}, \bibinfo {author} {\bibfnamefont {J.~C.}\ \bibnamefont {Pires}},
  \bibinfo {author} {\bibfnamefont {S.~W}\ \bibnamefont {Graham}}, \bibinfo
  {author} {\bibfnamefont {M.~A}\ \bibnamefont {McPherson}}, \bibinfo {author}
  {\bibfnamefont {L.~M}\ \bibnamefont {Prince}}, \bibinfo {author}
  {\bibfnamefont {T.~B}\ \bibnamefont {Patterson}}, \bibinfo {author}
  {\bibfnamefont {H.~S}\ \bibnamefont {Rai}}, \bibinfo {author} {\bibfnamefont
  {E.~H}\ \bibnamefont {Roalson}}, \bibinfo {author} {\bibfnamefont {T.~M}\
  \bibnamefont {Evans}}, \bibinfo {author} {\bibfnamefont {W.~J}\ \bibnamefont
  {Hahn}}, \bibinfo {author} {\bibfnamefont {K.~C}\ \bibnamefont {Millam}},
  \bibinfo {author} {\bibfnamefont {A.~W}\ \bibnamefont {Meerow}}, \bibinfo
  {author} {\bibfnamefont {M.}~\bibnamefont {Molvray}}, \bibinfo {author}
  {\bibfnamefont {P.~J}\ \bibnamefont {Kores}}, \bibinfo {author}
  {\bibfnamefont {H.~E}\ \bibnamefont {O'Brien}}, \bibinfo {author}
  {\bibfnamefont {J.~C}\ \bibnamefont {Hall}}, \bibinfo {author} {\bibfnamefont
  {W.~J.}\ \bibnamefont {Kress}}, \ and\ \bibinfo {author} {\bibfnamefont
  {K.~J}\ \bibnamefont {Sytsma}},\ }\bibfield  {title} {\enquote {\bibinfo
  {title} {{Repeated evolution of net venation and fleshy fruits among monocots
  in shaded habitats confirms a priori predictions: evidence from an ndhF
  phylogeny}},}\ }\href {\doibase 10.1098/rspb.2005.3067} {\bibfield  {journal}
  {\bibinfo  {journal} {Proceedings of the Royal Society B: Biological
  Sciences}\ }\textbf {\bibinfo {volume} {272}},\ \bibinfo {pages} {1481--1490}
  (\bibinfo {year} {2005})}\BibitemShut {NoStop}%
\bibitem [{\citenamefont {Blonder}\ \emph {et~al.}(2016)\citenamefont
  {Blonder}, \citenamefont {Baldwin}, \citenamefont {Enquist},\ and\
  \citenamefont {Robichaux}}]{Blonder2016}%
  \BibitemOpen
  \bibfield  {author} {\bibinfo {author} {\bibfnamefont {Benjamin}\
  \bibnamefont {Blonder}}, \bibinfo {author} {\bibfnamefont {Bruce~G.}\
  \bibnamefont {Baldwin}}, \bibinfo {author} {\bibfnamefont {Brian~J.}\
  \bibnamefont {Enquist}}, \ and\ \bibinfo {author} {\bibfnamefont {Robert~H.}\
  \bibnamefont {Robichaux}},\ }\bibfield  {title} {\enquote {\bibinfo {title}
  {{Variation and macroevolution in leaf functional traits in the Hawaiian
  silversword alliance (Asteraceae)}},}\ }\href {\doibase
  10.1111/1365-2745.12497} {\bibfield  {journal} {\bibinfo  {journal} {Journal
  of Ecology}\ }\textbf {\bibinfo {volume} {104}},\ \bibinfo {pages} {219--228}
  (\bibinfo {year} {2016})}\BibitemShut {NoStop}%
\bibitem [{\citenamefont {Steynen}\ and\ \citenamefont
  {Schultz}(2003)}]{Steynen2003}%
  \BibitemOpen
  \bibfield  {author} {\bibinfo {author} {\bibfnamefont {Quintin~J}\
  \bibnamefont {Steynen}}\ and\ \bibinfo {author} {\bibfnamefont {Elizabeth~A}\
  \bibnamefont {Schultz}},\ }\bibfield  {title} {\enquote {\bibinfo {title}
  {{The FORKED genes are essential for distal vein meeting in Arabidopsis.}}}\
  }\href {\doibase 10.1242/dev.00689} {\bibfield  {journal} {\bibinfo
  {journal} {Development (Cambridge, England)}\ }\textbf {\bibinfo {volume}
  {130}},\ \bibinfo {pages} {4695--4708} (\bibinfo {year} {2003})}\BibitemShut
  {NoStop}%
\bibitem [{\citenamefont {Carland}\ and\ \citenamefont
  {Nelson}(2009)}]{Carland2009}%
  \BibitemOpen
  \bibfield  {author} {\bibinfo {author} {\bibfnamefont {Francine}\
  \bibnamefont {Carland}}\ and\ \bibinfo {author} {\bibfnamefont {Timothy}\
  \bibnamefont {Nelson}},\ }\bibfield  {title} {\enquote {\bibinfo {title}
  {{CVP2- and CVL1-mediated phosphoinositide signaling as a regulator of the
  ARF GAP SFC/VAN3 in establishment of foliar vein patterns}},}\ }\href
  {\doibase 10.1111/j.1365-313X.2009.03920.x} {\bibfield  {journal} {\bibinfo
  {journal} {Plant Journal}\ }\textbf {\bibinfo {volume} {59}},\ \bibinfo
  {pages} {895--907} (\bibinfo {year} {2009})}\BibitemShut {NoStop}%
\bibitem [{\citenamefont {Berleth}\ \emph {et~al.}(2000)\citenamefont
  {Berleth}, \citenamefont {Mattsson},\ and\ \citenamefont
  {Hardtke}}]{Berleth2000}%
  \BibitemOpen
  \bibfield  {author} {\bibinfo {author} {\bibfnamefont {Thomas}\ \bibnamefont
  {Berleth}}, \bibinfo {author} {\bibfnamefont {Jim}\ \bibnamefont {Mattsson}},
  \ and\ \bibinfo {author} {\bibfnamefont {Christian~S.}\ \bibnamefont
  {Hardtke}},\ }\bibfield  {title} {\enquote {\bibinfo {title} {{Vascular
  continuity and auxin signals}},}\ }\href {\doibase
  10.1016/S1360-1385(00)01725-8} {\bibfield  {journal} {\bibinfo  {journal}
  {Trends in Plant Science}\ }\textbf {\bibinfo {volume} {5}},\ \bibinfo
  {pages} {387--393} (\bibinfo {year} {2000})}\BibitemShut {NoStop}%
\bibitem [{\citenamefont {le~Noble}\ \emph {et~al.}(2005)\citenamefont
  {le~Noble}, \citenamefont {Fleury}, \citenamefont {Pries}, \citenamefont
  {Corvol}, \citenamefont {Eichmann},\ and\ \citenamefont
  {Reneman}}]{LeNoble2005}%
  \BibitemOpen
  \bibfield  {author} {\bibinfo {author} {\bibfnamefont {F}~\bibnamefont
  {le~Noble}}, \bibinfo {author} {\bibfnamefont {V}~\bibnamefont {Fleury}},
  \bibinfo {author} {\bibfnamefont {A}~\bibnamefont {Pries}}, \bibinfo {author}
  {\bibfnamefont {P}~\bibnamefont {Corvol}}, \bibinfo {author} {\bibfnamefont
  {A}~\bibnamefont {Eichmann}}, \ and\ \bibinfo {author} {\bibfnamefont {R~S}\
  \bibnamefont {Reneman}},\ }\bibfield  {title} {\enquote {\bibinfo {title}
  {{Control of arterial branching morphogenesis in embryogenesis: go with the
  flow.}}}\ }\href {\doibase 10.1016/j.cardiores.2004.09.018} {\bibfield
  {journal} {\bibinfo  {journal} {Cardiovascular research}\ }\textbf {\bibinfo
  {volume} {65}},\ \bibinfo {pages} {619--28} (\bibinfo {year}
  {2005})}\BibitemShut {NoStop}%
\bibitem [{\citenamefont {Kurz}(2001)}]{Kurz2001}%
  \BibitemOpen
  \bibfield  {author} {\bibinfo {author} {\bibfnamefont {Haymo}\ \bibnamefont
  {Kurz}},\ }\bibfield  {title} {\enquote {\bibinfo {title} {{Physiology of
  angiogenesis.}}}\ }\href@noop {} {\bibfield  {journal} {\bibinfo  {journal}
  {Journal of Neuro-Oncology}\ }\textbf {\bibinfo {volume} {50}},\ \bibinfo
  {pages} {17--35} (\bibinfo {year} {2001})}\BibitemShut {NoStop}%
\bibitem [{\citenamefont {Nguyen}\ \emph {et~al.}(2006)\citenamefont {Nguyen},
  \citenamefont {Eichmann}, \citenamefont {{Le Noble}},\ and\ \citenamefont
  {Fleury}}]{Nguyen2006}%
  \BibitemOpen
  \bibfield  {author} {\bibinfo {author} {\bibfnamefont {Thi-Hanh}\
  \bibnamefont {Nguyen}}, \bibinfo {author} {\bibfnamefont {Anne}\ \bibnamefont
  {Eichmann}}, \bibinfo {author} {\bibfnamefont {Ferdinand}\ \bibnamefont {{Le
  Noble}}}, \ and\ \bibinfo {author} {\bibfnamefont {Vincent}\ \bibnamefont
  {Fleury}},\ }\bibfield  {title} {\enquote {\bibinfo {title} {{Dynamics of
  vascular branching morphogenesis: The effect of blood and tissue flow}},}\
  }\href {\doibase 10.1103/PhysRevE.73.061907} {\bibfield  {journal} {\bibinfo
  {journal} {Physical Review E}\ }\textbf {\bibinfo {volume} {73}},\ \bibinfo
  {pages} {061907} (\bibinfo {year} {2006})}\BibitemShut {NoStop}%
\bibitem [{\citenamefont {Ronellenfitsch}\ and\ \citenamefont
  {Katifori}(2016)}]{Ronellenfitsch2016}%
  \BibitemOpen
  \bibfield  {author} {\bibinfo {author} {\bibfnamefont {Henrik}\ \bibnamefont
  {Ronellenfitsch}}\ and\ \bibinfo {author} {\bibfnamefont {Eleni}\
  \bibnamefont {Katifori}},\ }\bibfield  {title} {\enquote {\bibinfo {title}
  {{Global Optimization, Local Adaptation, and the Role of Growth in
  Distribution Networks}},}\ }\href {\doibase 10.1103/PhysRevLett.117.138301}
  {\bibfield  {journal} {\bibinfo  {journal} {Physical Review Letters}\
  }\textbf {\bibinfo {volume} {117}},\ \bibinfo {pages} {138301} (\bibinfo
  {year} {2016})}\BibitemShut {NoStop}%
\bibitem [{\citenamefont {Smith}\ and\ \citenamefont
  {Bayer}(2009)}]{Smith2009}%
  \BibitemOpen
  \bibfield  {author} {\bibinfo {author} {\bibfnamefont {Richard~S.}\
  \bibnamefont {Smith}}\ and\ \bibinfo {author} {\bibfnamefont {Emmanuelle~M.}\
  \bibnamefont {Bayer}},\ }\bibfield  {title} {\enquote {\bibinfo {title}
  {{Auxin transport-feedback models of patterning in plants}},}\ }\href
  {\doibase 10.1111/j.1365-3040.2009.01997.x} {\bibfield  {journal} {\bibinfo
  {journal} {Plant, Cell {\&} Environment}\ }\textbf {\bibinfo {volume} {32}},\
  \bibinfo {pages} {1258--1271} (\bibinfo {year} {2009})}\BibitemShut {NoStop}%
\bibitem [{\citenamefont {Scarpella}(2006)}]{Scarpella2006}%
  \BibitemOpen
  \bibfield  {author} {\bibinfo {author} {\bibfnamefont {Enrico}\ \bibnamefont
  {Scarpella}},\ }\bibfield  {title} {\enquote {\bibinfo {title} {{Control of
  leaf vascular patterning by polar auxin transport}},}\ }\href {\doibase
  10.1101/gad.1402406} {\bibfield  {journal} {\bibinfo  {journal} {Genes {\&}
  Development}\ }\textbf {\bibinfo {volume} {20}},\ \bibinfo {pages}
  {1015--1027} (\bibinfo {year} {2006})}\BibitemShut {NoStop}%
\bibitem [{\citenamefont {Verna}\ \emph {et~al.}(2015)\citenamefont {Verna},
  \citenamefont {Sawchuk}, \citenamefont {Linh},\ and\ \citenamefont
  {Scarpella}}]{Verna2015}%
  \BibitemOpen
  \bibfield  {author} {\bibinfo {author} {\bibfnamefont {Carla}\ \bibnamefont
  {Verna}}, \bibinfo {author} {\bibfnamefont {Megan~G.}\ \bibnamefont
  {Sawchuk}}, \bibinfo {author} {\bibfnamefont {Nguyen~Manh}\ \bibnamefont
  {Linh}}, \ and\ \bibinfo {author} {\bibfnamefont {Enrico}\ \bibnamefont
  {Scarpella}},\ }\bibfield  {title} {\enquote {\bibinfo {title} {{Control of
  vein network topology by auxin transport}},}\ }\href {\doibase
  10.1186/s12915-015-0208-3} {\bibfield  {journal} {\bibinfo  {journal} {BMC
  Biology}\ }\textbf {\bibinfo {volume} {13}},\ \bibinfo {pages} {94} (\bibinfo
  {year} {2015})}\BibitemShut {NoStop}%
\bibitem [{\citenamefont {Feugier}\ \emph {et~al.}(2005)\citenamefont
  {Feugier}, \citenamefont {Mochizuki},\ and\ \citenamefont
  {Iwasa}}]{Feugier2005}%
  \BibitemOpen
  \bibfield  {author} {\bibinfo {author} {\bibfnamefont {Francois~G.}\
  \bibnamefont {Feugier}}, \bibinfo {author} {\bibfnamefont {A.}~\bibnamefont
  {Mochizuki}}, \ and\ \bibinfo {author} {\bibfnamefont {Y.}~\bibnamefont
  {Iwasa}},\ }\bibfield  {title} {\enquote {\bibinfo {title}
  {{Self-organization of the vascular system in plant leaves: Inter-dependent
  dynamics of auxin flux and carrier proteins}},}\ }\href {\doibase
  10.1016/j.jtbi.2005.03.017} {\bibfield  {journal} {\bibinfo  {journal}
  {Journal of Theoretical Biology}\ }\textbf {\bibinfo {volume} {236}},\
  \bibinfo {pages} {366--375} (\bibinfo {year} {2005})}\BibitemShut {NoStop}%
\bibitem [{\citenamefont {Feller}\ \emph {et~al.}(2015)\citenamefont {Feller},
  \citenamefont {Farcot},\ and\ \citenamefont {Mazza}}]{Feller2015}%
  \BibitemOpen
  \bibfield  {author} {\bibinfo {author} {\bibfnamefont {Chrystel}\
  \bibnamefont {Feller}}, \bibinfo {author} {\bibfnamefont {Etienne}\
  \bibnamefont {Farcot}}, \ and\ \bibinfo {author} {\bibfnamefont {Christian}\
  \bibnamefont {Mazza}},\ }\bibfield  {title} {\enquote {\bibinfo {title}
  {{Self-Organization of Plant Vascular Systems: Claims and Counter-Claims
  about the Flux-Based Auxin Transport Model.}}}\ }\href {\doibase
  10.1371/journal.pone.0118238} {\bibfield  {journal} {\bibinfo  {journal}
  {PloS one}\ }\textbf {\bibinfo {volume} {10}},\ \bibinfo {pages} {e0118238}
  (\bibinfo {year} {2015})}\BibitemShut {NoStop}%
\bibitem [{\citenamefont {Eichmann}\ \emph {et~al.}(2005)\citenamefont
  {Eichmann}, \citenamefont {Yuan}, \citenamefont {Moyon}, \citenamefont
  {Lenoble}, \citenamefont {Pardanaud},\ and\ \citenamefont
  {Breant}}]{Eichmann2005}%
  \BibitemOpen
  \bibfield  {author} {\bibinfo {author} {\bibfnamefont {Anne}\ \bibnamefont
  {Eichmann}}, \bibinfo {author} {\bibfnamefont {Li}~\bibnamefont {Yuan}},
  \bibinfo {author} {\bibfnamefont {Delphine}\ \bibnamefont {Moyon}}, \bibinfo
  {author} {\bibfnamefont {Ferdinand}\ \bibnamefont {Lenoble}}, \bibinfo
  {author} {\bibfnamefont {Luc}\ \bibnamefont {Pardanaud}}, \ and\ \bibinfo
  {author} {\bibfnamefont {Christiane}\ \bibnamefont {Breant}},\ }\bibfield
  {title} {\enquote {\bibinfo {title} {{Vascular development: from precursor
  cells to branched arterial and venous networks}},}\ }\href {\doibase
  10.1387/ijdb.041941ae} {\bibfield  {journal} {\bibinfo  {journal} {The
  International Journal of Developmental Biology}\ }\textbf {\bibinfo {volume}
  {49}},\ \bibinfo {pages} {259--267} (\bibinfo {year} {2005})}\BibitemShut
  {NoStop}%
\bibitem [{\citenamefont {Hu}\ \emph {et~al.}(2012)\citenamefont {Hu},
  \citenamefont {Cai},\ and\ \citenamefont {Rangan}}]{Hu2012}%
  \BibitemOpen
  \bibfield  {author} {\bibinfo {author} {\bibfnamefont {Dan}\ \bibnamefont
  {Hu}}, \bibinfo {author} {\bibfnamefont {David}\ \bibnamefont {Cai}}, \ and\
  \bibinfo {author} {\bibfnamefont {Aaditya~V}\ \bibnamefont {Rangan}},\
  }\bibfield  {title} {\enquote {\bibinfo {title} {{Blood vessel adaptation
  with fluctuations in capillary flow distribution.}}}\ }\href {\doibase
  10.1371/journal.pone.0045444} {\bibfield  {journal} {\bibinfo  {journal}
  {PloS one}\ }\textbf {\bibinfo {volume} {7}},\ \bibinfo {pages} {e45444}
  (\bibinfo {year} {2012})}\BibitemShut {NoStop}%
\bibitem [{\citenamefont {Scianna}\ \emph {et~al.}(2013)\citenamefont
  {Scianna}, \citenamefont {Bell},\ and\ \citenamefont
  {Preziosi}}]{Scianna2013}%
  \BibitemOpen
  \bibfield  {author} {\bibinfo {author} {\bibfnamefont {M}~\bibnamefont
  {Scianna}}, \bibinfo {author} {\bibfnamefont {C~G}\ \bibnamefont {Bell}}, \
  and\ \bibinfo {author} {\bibfnamefont {L}~\bibnamefont {Preziosi}},\
  }\bibfield  {title} {\enquote {\bibinfo {title} {{A review of mathematical
  models for the formation of vascular networks.}}}\ }\href {\doibase
  10.1016/j.jtbi.2013.04.037} {\bibfield  {journal} {\bibinfo  {journal}
  {Journal of theoretical biology}\ }\textbf {\bibinfo {volume} {333}},\
  \bibinfo {pages} {174--209} (\bibinfo {year} {2013})}\BibitemShut {NoStop}%
\bibitem [{\citenamefont {Hacking}\ \emph {et~al.}(1996)\citenamefont
  {Hacking}, \citenamefont {VanBavel},\ and\ \citenamefont
  {Spaan}}]{Hacking1996}%
  \BibitemOpen
  \bibfield  {author} {\bibinfo {author} {\bibfnamefont {W~J}\ \bibnamefont
  {Hacking}}, \bibinfo {author} {\bibfnamefont {E}~\bibnamefont {VanBavel}}, \
  and\ \bibinfo {author} {\bibfnamefont {J~A~E}\ \bibnamefont {Spaan}},\
  }\bibfield  {title} {\enquote {\bibinfo {title} {{Shear stress is not
  sufficient to control growth of vascular networks: a model study}},}\ }\href
  {http://www.ncbi.nlm.nih.gov/pubmed/8769773} {\bibfield  {journal} {\bibinfo
  {journal} {The American Journal of Physiology}\ }\textbf {\bibinfo {volume}
  {270}},\ \bibinfo {pages} {H364--75} (\bibinfo {year} {1996})}\BibitemShut
  {NoStop}%
\bibitem [{\citenamefont {Chen}\ \emph {et~al.}(2012)\citenamefont {Chen},
  \citenamefont {Jiang}, \citenamefont {Li}, \citenamefont {Hu}, \citenamefont
  {Bu}, \citenamefont {Cai},\ and\ \citenamefont {Du}}]{Chen2012}%
  \BibitemOpen
  \bibfield  {author} {\bibinfo {author} {\bibfnamefont {Qi}~\bibnamefont
  {Chen}}, \bibinfo {author} {\bibfnamefont {Luan}\ \bibnamefont {Jiang}},
  \bibinfo {author} {\bibfnamefont {Chun}\ \bibnamefont {Li}}, \bibinfo
  {author} {\bibfnamefont {Dan}\ \bibnamefont {Hu}}, \bibinfo {author}
  {\bibfnamefont {Ji-wen}\ \bibnamefont {Bu}}, \bibinfo {author} {\bibfnamefont
  {David}\ \bibnamefont {Cai}}, \ and\ \bibinfo {author} {\bibfnamefont
  {Jiu-lin}\ \bibnamefont {Du}},\ }\bibfield  {title} {\enquote {\bibinfo
  {title} {{Haemodynamics-driven developmental pruning of brain vasculature in
  zebrafish.}}}\ }\href {\doibase 10.1371/journal.pbio.1001374} {\bibfield
  {journal} {\bibinfo  {journal} {PLOS Biology}\ }\textbf {\bibinfo {volume}
  {10}},\ \bibinfo {pages} {e1001374} (\bibinfo {year} {2012})}\BibitemShut
  {NoStop}%
\bibitem [{\citenamefont {Marcos}\ and\ \citenamefont
  {Berleth}(2014)}]{Marcos2014}%
  \BibitemOpen
  \bibfield  {author} {\bibinfo {author} {\bibfnamefont {Danielle}\
  \bibnamefont {Marcos}}\ and\ \bibinfo {author} {\bibfnamefont {Thomas}\
  \bibnamefont {Berleth}},\ }\bibfield  {title} {\enquote {\bibinfo {title}
  {{Dynamic auxin transport patterns preceding vein formation revealed by
  live-imaging of Arabidopsis leaf primordia}},}\ }\href {\doibase
  10.3389/fpls.2014.00235} {\bibfield  {journal} {\bibinfo  {journal}
  {Frontiers in Plant Science}\ }\textbf {\bibinfo {volume} {5}},\ \bibinfo
  {pages} {235} (\bibinfo {year} {2014})}\BibitemShut {NoStop}%
\bibitem [{\citenamefont {Hu}\ and\ \citenamefont {Cai}(2013)}]{Hu2013}%
  \BibitemOpen
  \bibfield  {author} {\bibinfo {author} {\bibfnamefont {Dan}\ \bibnamefont
  {Hu}}\ and\ \bibinfo {author} {\bibfnamefont {David}\ \bibnamefont {Cai}},\
  }\bibfield  {title} {\enquote {\bibinfo {title} {{Adaptation and Optimization
  of Biological Transport Networks}},}\ }\href {\doibase
  10.1103/PhysRevLett.111.138701} {\bibfield  {journal} {\bibinfo  {journal}
  {Physical Review Letters}\ }\textbf {\bibinfo {volume} {111}},\ \bibinfo
  {pages} {138701} (\bibinfo {year} {2013})}\BibitemShut {NoStop}%
\bibitem [{\citenamefont {Rolland-Lagan}\ and\ \citenamefont
  {Prusinkiewicz}(2005)}]{Rolland-Lagan2005}%
  \BibitemOpen
  \bibfield  {author} {\bibinfo {author} {\bibfnamefont {Anne-Ga{\"{e}}lle}\
  \bibnamefont {Rolland-Lagan}}\ and\ \bibinfo {author} {\bibfnamefont
  {Przemyslaw}\ \bibnamefont {Prusinkiewicz}},\ }\bibfield  {title} {\enquote
  {\bibinfo {title} {{Reviewing models of auxin canalization in the context of
  leaf vein pattern formation in Arabidopsis}},}\ }\href {\doibase
  10.1111/j.1365-313X.2005.02581.x} {\bibfield  {journal} {\bibinfo  {journal}
  {The Plant Journal}\ }\textbf {\bibinfo {volume} {44}},\ \bibinfo {pages}
  {854--865} (\bibinfo {year} {2005})}\BibitemShut {NoStop}%
\bibitem [{\citenamefont {van Berkel}\ \emph {et~al.}(2013)\citenamefont {van
  Berkel}, \citenamefont {de~Boer}, \citenamefont {Scheres},\ and\
  \citenamefont {ten Tusscher}}]{VanBerkel2013}%
  \BibitemOpen
  \bibfield  {author} {\bibinfo {author} {\bibfnamefont {Klaartje}\
  \bibnamefont {van Berkel}}, \bibinfo {author} {\bibfnamefont {Rob~J}\
  \bibnamefont {de~Boer}}, \bibinfo {author} {\bibfnamefont {Ben}\ \bibnamefont
  {Scheres}}, \ and\ \bibinfo {author} {\bibfnamefont {Kirsten}\ \bibnamefont
  {ten Tusscher}},\ }\bibfield  {title} {\enquote {\bibinfo {title} {{Polar
  auxin transport: models and mechanisms}},}\ }\href {\doibase
  10.1242/dev.079111} {\bibfield  {journal} {\bibinfo  {journal} {Development}\
  }\textbf {\bibinfo {volume} {140}},\ \bibinfo {pages} {2253--2268} (\bibinfo
  {year} {2013})}\BibitemShut {NoStop}%
\bibitem [{Note1()}]{Note1}%
  \BibitemOpen
  \bibinfo {note} {See Supplemental Material [url], which includes Refs.~\cite
  {Blakeslee2005,Kramer2006}.}\BibitemShut {Stop}%
\bibitem [{\citenamefont {Fleury}\ and\ \citenamefont
  {Unbekandt}(2007)}]{Fleury2007}%
  \BibitemOpen
  \bibfield  {author} {\bibinfo {author} {\bibfnamefont {Vincent}\ \bibnamefont
  {Fleury}}\ and\ \bibinfo {author} {\bibfnamefont {Mathieu}\ \bibnamefont
  {Unbekandt}},\ }\bibfield  {title} {\enquote {\bibinfo {title} {{The Textural
  Aspects of Vessel Formation during Embryo Development and Their Relation to
  Gastrulation Movements}},}\ }\href@noop {} {\bibfield  {journal} {\bibinfo
  {journal} {Organogenesis}\ }\textbf {\bibinfo {volume} {3}},\ \bibinfo
  {pages} {49--56} (\bibinfo {year} {2007})}\BibitemShut {NoStop}%
\bibitem [{\citenamefont {Fruttiger}(2007)}]{Fruttiger2007}%
  \BibitemOpen
  \bibfield  {author} {\bibinfo {author} {\bibfnamefont {Marcus}\ \bibnamefont
  {Fruttiger}},\ }\bibfield  {title} {\enquote {\bibinfo {title} {{Development
  of the retinal vasculature}},}\ }\href {\doibase 10.1007/s10456-007-9065-1}
  {\bibfield  {journal} {\bibinfo  {journal} {Angiogenesis}\ }\textbf {\bibinfo
  {volume} {10}},\ \bibinfo {pages} {77--88} (\bibinfo {year}
  {2007})}\BibitemShut {NoStop}%
\bibitem [{\citenamefont {Bernot}\ \emph {et~al.}(2009)\citenamefont {Bernot},
  \citenamefont {Caselles},\ and\ \citenamefont {Morel}}]{Bernot2009}%
  \BibitemOpen
  \bibfield  {author} {\bibinfo {author} {\bibfnamefont {Marc}\ \bibnamefont
  {Bernot}}, \bibinfo {author} {\bibfnamefont {Vincent}\ \bibnamefont
  {Caselles}}, \ and\ \bibinfo {author} {\bibfnamefont {Jean-Michel}\
  \bibnamefont {Morel}},\ }\href {\doibase 10.1007/978-3-540-69315-4} {\emph
  {\bibinfo {title} {{Optimal Transportation Networks}}}},\ \bibinfo {series}
  {Lecture Notes in Mathematics}, Vol.\ \bibinfo {volume} {1955}\ (\bibinfo
  {publisher} {Springer Berlin Heidelberg},\ \bibinfo {address} {Berlin,
  Heidelberg},\ \bibinfo {year} {2009})\BibitemShut {NoStop}%
\bibitem [{\citenamefont {Banavar}\ \emph {et~al.}(2000)\citenamefont
  {Banavar}, \citenamefont {Colaiori}, \citenamefont {Flammini}, \citenamefont
  {Maritan},\ and\ \citenamefont {Rinaldo}}]{Banavar2000}%
  \BibitemOpen
  \bibfield  {author} {\bibinfo {author} {\bibfnamefont {Jayanth~R}\
  \bibnamefont {Banavar}}, \bibinfo {author} {\bibfnamefont {Francesca}\
  \bibnamefont {Colaiori}}, \bibinfo {author} {\bibfnamefont {Alessandro}\
  \bibnamefont {Flammini}}, \bibinfo {author} {\bibfnamefont {Amos}\
  \bibnamefont {Maritan}}, \ and\ \bibinfo {author} {\bibfnamefont {Andrea}\
  \bibnamefont {Rinaldo}},\ }\bibfield  {title} {\enquote {\bibinfo {title}
  {{Topology of the Fittest Transportation Network}},}\ }\href
  {file:///Users/eleni/Library/Application
  Support/Papers2/Files/Banavar2000.pdf
  papers2://publication/doi/10.1103/PhysRevLett.84.4745} {\bibfield  {journal}
  {\bibinfo  {journal} {Physical Review Letters}\ }\textbf {\bibinfo {volume}
  {84}},\ \bibinfo {pages} {4745--4748} (\bibinfo {year} {2000})}\BibitemShut
  {NoStop}%
\bibitem [{\citenamefont {Feugier}\ and\ \citenamefont
  {Iwasa}(2006)}]{Feugier2006}%
  \BibitemOpen
  \bibfield  {author} {\bibinfo {author} {\bibfnamefont {Fran{\c{c}}ois~G.}\
  \bibnamefont {Feugier}}\ and\ \bibinfo {author} {\bibfnamefont {Yoh}\
  \bibnamefont {Iwasa}},\ }\bibfield  {title} {\enquote {\bibinfo {title} {{How
  canalization can make loops: A new model of reticulated leaf vascular pattern
  formation}},}\ }\href {\doibase 10.1016/j.jtbi.2006.05.022} {\bibfield
  {journal} {\bibinfo  {journal} {Journal of Theoretical Biology}\ }\textbf
  {\bibinfo {volume} {243}},\ \bibinfo {pages} {235--244} (\bibinfo {year}
  {2006})}\BibitemShut {NoStop}%
\bibitem [{\citenamefont {Drew}\ \emph {et~al.}(2011)\citenamefont {Drew},
  \citenamefont {Shih},\ and\ \citenamefont {Kleinfeld}}]{Drew2011}%
  \BibitemOpen
  \bibfield  {author} {\bibinfo {author} {\bibfnamefont {Patrick~J}\
  \bibnamefont {Drew}}, \bibinfo {author} {\bibfnamefont {Andy~Y}\ \bibnamefont
  {Shih}}, \ and\ \bibinfo {author} {\bibfnamefont {David}\ \bibnamefont
  {Kleinfeld}},\ }\bibfield  {title} {\enquote {\bibinfo {title} {{Fluctuating
  and sensory-induced vasodynamics in rodent cortex extend arteriole
  capacity.}}}\ }\href {\doibase 10.1073/pnas.1100428108} {\bibfield  {journal}
  {\bibinfo  {journal} {Proceedings of the National Academy of Sciences of the
  United States of America}\ }\textbf {\bibinfo {volume} {108}},\ \bibinfo
  {pages} {8473--8478} (\bibinfo {year} {2011})}\BibitemShut {NoStop}%
\bibitem [{\citenamefont {Corson}(2010)}]{Corson2010}%
  \BibitemOpen
  \bibfield  {author} {\bibinfo {author} {\bibfnamefont {Francis}\ \bibnamefont
  {Corson}},\ }\bibfield  {title} {\enquote {\bibinfo {title} {{Fluctuations
  and Redundancy in Optimal Transport Networks}},}\ }\href {\doibase
  10.1103/PhysRevLett.104.048703} {\bibfield  {journal} {\bibinfo  {journal}
  {Physical Review Letters}\ }\textbf {\bibinfo {volume} {104}},\ \bibinfo
  {pages} {048703} (\bibinfo {year} {2010})},\ \Eprint
  {http://arxiv.org/abs/0905.4947} {arXiv:0905.4947} \BibitemShut {NoStop}%
\bibitem [{\citenamefont {Katifori}\ \emph {et~al.}(2010)\citenamefont
  {Katifori}, \citenamefont {Sz{\"{o}}ll{\H{o}}si},\ and\ \citenamefont
  {Magnasco}}]{Katifori2010}%
  \BibitemOpen
  \bibfield  {author} {\bibinfo {author} {\bibfnamefont {Eleni}\ \bibnamefont
  {Katifori}}, \bibinfo {author} {\bibfnamefont {Gergely~J.}\ \bibnamefont
  {Sz{\"{o}}ll{\H{o}}si}}, \ and\ \bibinfo {author} {\bibfnamefont
  {Marcelo~O.}\ \bibnamefont {Magnasco}},\ }\bibfield  {title} {\enquote
  {\bibinfo {title} {{Damage and Fluctuations Induce Loops in Optimal Transport
  Networks}},}\ }\href {\doibase 10.1103/PhysRevLett.104.048704} {\bibfield
  {journal} {\bibinfo  {journal} {Physical Review Letters}\ }\textbf {\bibinfo
  {volume} {104}},\ \bibinfo {pages} {048704} (\bibinfo {year}
  {2010})}\BibitemShut {NoStop}%
\bibitem [{\citenamefont {Ronellenfitsch}\ \emph {et~al.}(2018)\citenamefont
  {Ronellenfitsch}, \citenamefont {Dunkel},\ and\ \citenamefont
  {Wilczek}}]{Ronellenfitsch2018a}%
  \BibitemOpen
  \bibfield  {author} {\bibinfo {author} {\bibfnamefont {Henrik}\ \bibnamefont
  {Ronellenfitsch}}, \bibinfo {author} {\bibfnamefont {J{\"{o}}rn}\
  \bibnamefont {Dunkel}}, \ and\ \bibinfo {author} {\bibfnamefont {Michael}\
  \bibnamefont {Wilczek}},\ }\bibfield  {title} {\enquote {\bibinfo {title}
  {{Optimal Noise-Canceling Networks}},}\ }\href {\doibase
  10.1103/PhysRevLett.121.208301} {\bibfield  {journal} {\bibinfo  {journal}
  {Physical Review Letters}\ }\textbf {\bibinfo {volume} {121}},\ \bibinfo
  {pages} {208301} (\bibinfo {year} {2018})},\ \Eprint
  {http://arxiv.org/abs/1807.08376v2} {arXiv:1807.08376v2} \BibitemShut
  {NoStop}%
\bibitem [{\citenamefont {Gr{\"a}wer}\ \emph {et~al.}(2015)\citenamefont
  {Gr{\"a}wer}, \citenamefont {Modes}, \citenamefont {Magnasco},\ and\
  \citenamefont {Katifori}}]{Graewer2015}%
  \BibitemOpen
  \bibfield  {author} {\bibinfo {author} {\bibfnamefont {Johannes}\
  \bibnamefont {Gr{\"a}wer}}, \bibinfo {author} {\bibfnamefont {Carl~D.}\
  \bibnamefont {Modes}}, \bibinfo {author} {\bibfnamefont {Marcelo~O.}\
  \bibnamefont {Magnasco}}, \ and\ \bibinfo {author} {\bibfnamefont {Eleni}\
  \bibnamefont {Katifori}},\ }\bibfield  {title} {\enquote {\bibinfo {title}
  {{Structural self-assembly and avalanchelike dynamics in locally adaptive
  networks}},}\ }\href {\doibase 10.1103/PhysRevE.92.012801} {\bibfield
  {journal} {\bibinfo  {journal} {Physical Review E}\ }\textbf {\bibinfo
  {volume} {92}},\ \bibinfo {pages} {012801} (\bibinfo {year} {2015})},\
  \Eprint {http://arxiv.org/abs/1405.7870} {arXiv:1405.7870} \BibitemShut
  {NoStop}%
\bibitem [{\citenamefont {Martens}\ and\ \citenamefont
  {Klemm}(2017)}]{Martens2017}%
  \BibitemOpen
  \bibfield  {author} {\bibinfo {author} {\bibfnamefont {Erik~Andreas}\
  \bibnamefont {Martens}}\ and\ \bibinfo {author} {\bibfnamefont {Konstantin}\
  \bibnamefont {Klemm}},\ }\bibfield  {title} {\enquote {\bibinfo {title}
  {{Transitions from trees to cycles in adaptive flow networks}},}\ }\href
  {\doibase 10.3389/fphy.2017.00062} {\bibfield  {journal} {\bibinfo  {journal}
  {Front. Phys.}\ }\textbf {\bibinfo {volume} {5}},\ \bibinfo {pages} {1--10}
  (\bibinfo {year} {2017})},\ \Eprint {http://arxiv.org/abs/1711.00401}
  {arXiv:1711.00401} \BibitemShut {NoStop}%
\bibitem [{\citenamefont {Fiorin}\ \emph {et~al.}(2015)\citenamefont {Fiorin},
  \citenamefont {Brodribb},\ and\ \citenamefont {Anfodillo}}]{Fiorin2015}%
  \BibitemOpen
  \bibfield  {author} {\bibinfo {author} {\bibfnamefont {Lucia}\ \bibnamefont
  {Fiorin}}, \bibinfo {author} {\bibfnamefont {Timothy~J}\ \bibnamefont
  {Brodribb}}, \ and\ \bibinfo {author} {\bibfnamefont {Tommaso}\ \bibnamefont
  {Anfodillo}},\ }\bibfield  {title} {\enquote {\bibinfo {title} {{Transport
  efficiency through uniformity: organization of veins and stomata in
  angiosperm leaves}},}\ }\href {\doibase 10.1111/nph.13577} {\bibfield
  {journal} {\bibinfo  {journal} {New Phytologist}\ }\textbf {\bibinfo {volume}
  {209}},\ \bibinfo {pages} {216--227} (\bibinfo {year} {2015})}\BibitemShut
  {NoStop}%
\bibitem [{\citenamefont {Geilen}\ and\ \citenamefont
  {Basten}(2007)}]{Geilen2007}%
  \BibitemOpen
  \bibfield  {author} {\bibinfo {author} {\bibfnamefont {Marc}\ \bibnamefont
  {Geilen}}\ and\ \bibinfo {author} {\bibfnamefont {Twan}\ \bibnamefont
  {Basten}},\ }\bibfield  {title} {\enquote {\bibinfo {title} {{A Calculator
  for Pareto Points}},}\ }in\ \href {\doibase 10.1109/DATE.2007.364605} {\emph
  {\bibinfo {booktitle} {2007 Design, Automation {\&} Test in Europe Conference
  {\&} Exhibition}}},\ Vol.~\bibinfo {volume} {2}\ (\bibinfo  {publisher}
  {IEEE},\ \bibinfo {year} {2007})\ pp.\ \bibinfo {pages} {1--6}\BibitemShut
  {NoStop}%
\bibitem [{\citenamefont {Zhou}\ and\ \citenamefont {Zheng}(2003)}]{Zhou2003}%
  \BibitemOpen
  \bibfield  {author} {\bibinfo {author} {\bibfnamefont {Zhiyan}\ \bibnamefont
  {Zhou}}\ and\ \bibinfo {author} {\bibfnamefont {Shaolin}\ \bibnamefont
  {Zheng}},\ }\bibfield  {title} {\enquote {\bibinfo {title} {{The missing link
  in Ginkgo evolution}},}\ }\href {\doibase 10.1038/423821a} {\bibfield
  {journal} {\bibinfo  {journal} {Nature}\ }\textbf {\bibinfo {volume} {423}},\
  \bibinfo {pages} {821--822} (\bibinfo {year} {2003})}\BibitemShut {NoStop}%
\bibitem [{\citenamefont {D{\"{o}}rken}(2014)}]{Dorken2014}%
  \BibitemOpen
  \bibfield  {author} {\bibinfo {author} {\bibfnamefont {Veit~Martin}\
  \bibnamefont {D{\"{o}}rken}},\ }\bibfield  {title} {\enquote {\bibinfo
  {title} {{Morphology, anatomy and vasculature in leaves of Ginkgo biloba L.
  (Ginkgoaceae, Ginkgoales) under functional and evolutionary aspects}},}\
  }\href {\doibase 10.1002/fedr.201400008} {\bibfield  {journal} {\bibinfo
  {journal} {Feddes Repertorium}\ }\textbf {\bibinfo {volume} {124}},\ \bibinfo
  {pages} {80--97} (\bibinfo {year} {2014})}\BibitemShut {NoStop}%
\bibitem [{\citenamefont {Blakeslee}\ \emph {et~al.}(2005)\citenamefont
  {Blakeslee}, \citenamefont {Peer},\ and\ \citenamefont
  {Murphy}}]{Blakeslee2005}%
  \BibitemOpen
  \bibfield  {author} {\bibinfo {author} {\bibfnamefont {Joshua~J.}\
  \bibnamefont {Blakeslee}}, \bibinfo {author} {\bibfnamefont {Wendy~A.}\
  \bibnamefont {Peer}}, \ and\ \bibinfo {author} {\bibfnamefont {Angus~S.}\
  \bibnamefont {Murphy}},\ }\bibfield  {title} {\enquote {\bibinfo {title}
  {{Auxin transport}},}\ }\href {\doibase 10.1016/j.pbi.2005.07.014} {\bibfield
   {journal} {\bibinfo  {journal} {Current Opinion in Plant Biology}\ }\textbf
  {\bibinfo {volume} {8}},\ \bibinfo {pages} {494--500} (\bibinfo {year}
  {2005})}\BibitemShut {NoStop}%
\bibitem [{\citenamefont {Kramer}\ and\ \citenamefont
  {Bennett}(2006)}]{Kramer2006}%
  \BibitemOpen
  \bibfield  {author} {\bibinfo {author} {\bibfnamefont {Eric~M.}\ \bibnamefont
  {Kramer}}\ and\ \bibinfo {author} {\bibfnamefont {Malcolm~J.}\ \bibnamefont
  {Bennett}},\ }\bibfield  {title} {\enquote {\bibinfo {title} {{Auxin
  transport: a field in flux}},}\ }\href {\doibase
  10.1016/j.tplants.2006.06.002} {\bibfield  {journal} {\bibinfo  {journal}
  {Trends in Plant Science}\ }\textbf {\bibinfo {volume} {11}},\ \bibinfo
  {pages} {382--386} (\bibinfo {year} {2006})}\BibitemShut {NoStop}%
\end{thebibliography}
\end{document}